\documentclass[12pt, authoryear]{article}
\usepackage{amsfonts, graphicx,color,multirow, amsthm,amsmath,natbib}
\usepackage{setspace,url}
\usepackage{multirow}
\usepackage{bm}
\def\boxit#1{\vbox{\hrule\hbox{\vrule\kern6pt\vbox{\kern6pt#1\kern6pt}\kern6pt\vrule}\hrule}}

\usepackage[colorlinks,citecolor=blue,urlcolor=blue]{hyperref}
\allowdisplaybreaks
\usepackage{enumitem}
\usepackage[margin=1in]{geometry}

\begin{document}

\newcommand{\argmax}[0]{\mbox{argmax}}
\newcommand{\argmin}[0]{\mbox{argmin}}
\newcommand{\ol}[1]{\overline{#1}}

\theoremstyle{plain}
\newtheorem{thm}{Theorem}
\newtheorem{lem}{Lemma}
\newtheorem{prop}[thm]{Proposition}
\newtheorem{cor}[thm]{Corollary}

\theoremstyle{definition}
\newtheorem{defn}{Definition}[section]
\newtheorem{conj}{Conjecture}[section]
\newtheorem{exmp}{Example}[section]
\newtheorem{condition}{Condition}[section]

\theoremstyle{remark}
\newtheorem{rmk}{Remark}
\newtheorem{note}{Note}
\newtheorem{case}{Case}

\newcommand{\var}{\mbox{var}}
\newcommand{\Var}{\mbox{Var}}
\newcommand{\bb}{\mbox{\bf b}}
\newcommand{\bff}{\mbox{\bf f}}
\newcommand{\bx}{\mbox{\bf x}}
\newcommand{\by}{\mbox{\bf y}}
\newcommand{\bg}{\mbox{\bf g}}
\newcommand{\bA}{\mbox{\bf A}}
\newcommand{\ba}{\mbox{\bf a}}
\newcommand{\bB}{\mbox{\bf B}}
\newcommand{\bC}{\mbox{\bf C}}
\newcommand{\bD}{\mbox{\bf D}}
\newcommand{\bE}{\mbox{\bf E}}
\newcommand{\bF}{\mbox{\bf F}}
\newcommand{\bG}{\mbox{\bf G}}
\newcommand{\bH}{\mbox{\bf H}}
\newcommand{\bI}{\mbox{\bf I}}
\newcommand{\bU}{\mbox{\bf U}}
\newcommand{\bV}{\mbox{\bf V}}
\newcommand{\bQ}{\mbox{\bf Q}}
\newcommand{\bR}{\mbox{\bf R}}
\newcommand{\bS}{\mbox{\bf S}}
\newcommand{\bX}{\mbox{\bf X}}
\newcommand{\bY}{\mbox{\bf Y}}
\newcommand{\bZ}{\mbox{\bf Z}}
\newcommand{\br}{\mbox{\bf r}}
\newcommand{\bv}{\mbox{\bf v}}
\newcommand{\bL}{\mbox{\bf L}}
\newcommand{\bbP}{\mathbb{P} }
\newcommand{\bone}{\mbox{\bf 1}}
\newcommand{\bsone}{\mbox{\scriptsize \bf 1}}
\newcommand{\bzero}{\mbox{\bf 0}}
\newcommand{\bveps}{\mbox{\boldmath $\varepsilon$}}
\newcommand{\bet}{\mbox{\boldmath $\eta$}}
\newcommand{\bxi}{\mbox{\boldmath $\xi$}}
\newcommand{\bzeta}{\mbox{\boldmath $\zeta$}}
\newcommand{\beps}{\mbox{\boldmath $\varepsilon$}}
\newcommand{\bbeta}{\mbox{\boldmath $\beta$}}
\newcommand{\balpha}{\mbox{\boldmath $\alpha$}}
\newcommand{\bPsi}{\mbox{\boldmath $\Psi$}}
\newcommand{\bomega}{\mbox{\boldmath $\omega$}}
\newcommand{\hbbeta}{\hat{\bbeta}}
\newcommand{\hbeta}{\hat{\beta}}
\newcommand{\Bbeta}{ \bar{\beta}}
\newcommand{\bmu}{\mbox{\boldmath $\mu$}}
\newcommand{\bgamma}{\mbox{\boldmath $\gamma$}}
\newcommand{\mv}{\mbox{V}}
\newcommand{\bSigma}{\mbox{\boldmath $\Sigma$}}
\newcommand{\cov}{\mbox{cov}}
\newcommand{\beq}{\begin{eqnarray}}
\newcommand{\eeq}{\end{eqnarray}}
\newcommand{\beqn}{\begin{eqnarray*}}
\newcommand{\eeqn}{\end{eqnarray*}}
\newcommand{\Mstar}{{\mathcal{M}_{\star}}}
\newcommand{\lammax}{{\lambda_{\max}}}
\newcommand{\A}{{\bbP_n \bPsi_jY}}
\newcommand{\B}{{(\bbP_n \bPsi_j\bPsi_j^T)^{-1}}}
\newcommand{\ea}{{E \bPsi_jY}}
\newcommand{\eb}{{(E \bPsi_j\bPsi_j^T)^{-1}}}

\title{\bf Partial Distance Correlation Screening for High Dimensional Time Series\date{}
\thanks{Kashif Yousuf, the corresponding author, is PhD Candidate (Email: ky2304@columbia.edu) and Yang Feng is Associate Professor (Email: yangfeng@stat.columbia.edu), Department of Statistics, Columbia University, New York, NY 10027. Yang Feng is partially supported by NSF Grant DMS-1554804.
}}

\author{Kashif Yousuf and Yang Feng}
\maketitle


\begin{abstract}
\begin{singlespace}
High dimensional time series datasets are becoming increasingly common in various fields such as economics, finance, meteorology, and neuroscience. Given this ubiquity of time series data, it is surprising that very few works on variable screening discuss the time series setting, and even fewer works have developed methods which utilize the unique features of time series data. This paper introduces several model free screening methods based on the partial distance correlation and developed specifically to deal with time dependent data. Methods are developed both for univariate models, such as nonlinear autoregressive models with exogenous predictors (NARX), and multivariate models such as linear or nonlinear VAR models. Sure screening properties are proved for our methods, which depend on the moment conditions, and the strength of dependence in the response and covariate processes, amongst other factors. Dependence is quantified by functional dependence measures \citep{Wu2005} and $\beta$-mixing coefficients, and the results rely on the use of Nagaev and Rosenthal type inequalities for dependent random variables. Finite sample performance of our methods is shown through extensive simulation studies, and we include an application to macroeconomic forecasting.
\end{singlespace}
\end{abstract}


\clearpage

\section{Introduction} \label{sec:introduction}

High dimensionality is an increasingly common characteristic of data being collected in fields as diverse as genetics, neuroscience, astronomy, finance, and macroeconomics. In these fields, we frequently encounter situations in which the number of candidate predictors ($p_n$) is much larger than the number of samples ($n$), and statistical inference is made possible by relying on the assumption of sparsity. The sparsity assumption, which states that only a small number of covariates contributes to the response, has led to a wealth of theoretical results and methods available for identifying important predictors in this high dimensional setting. These methods broadly fall into two classes: screening methods and penalized likelihood methods, and we focus on the screening approach in this work. For the case where $p_n$ is much larger than $n$, screening is more computationally feasible as a first stage method, which can be followed by a second stage method, such as penalized likelihood approaches, on the reduced subset of predictors selected at the screening stage.

\cite{FanLv2008} proposed Sure Independence Screening (SIS) for the linear model, and it is based on ranking the magnitudes of the marginal Pearson correlations of the covariates with the response. A large amount of work has been done since then to generalize the procedure to various other types of models including: generalized linear models \citep{FanSong2010}, nonparametric additive models \citep{Feng2011}, Cox proportional hazards model \citep{fan2010high}, linear quantile models \citep{Ma2017}, and varying coefficient models \citep{Fanetal2014}. Model-free screening methods, which do not assume any particular model \textit{a priori}, have also been developed. Some examples include: a distance correlation based method  in \cite{Lietal2012}, the fused Kolmogorov filter in \cite{mai2015fused}, a conditional distance correlation method in \cite{liu2017}, a method based on maximum correlation in \cite{Huang2016}, a martingale difference based approach in \cite{Shao2014}, and a smoothing bandwidth based method in \cite{feng2017nonparametric}. For a partial survey of screening methods, one can consult \cite{Liuetal2015}. The main theoretical result of these methods is the so called ``sure screening property", which states that under appropriate conditions we can reduce the dimension of the feature space from size $p_n=O\left(\exp\left(n^\alpha\right)\right)$ to a far smaller size $d_n$, while retaining all the relevant predictors with probability approaching 1.

Although there has been a large amount of interest in developing screening methods, it is surprising to see that almost all of the works operate under the assumption of independent observations. This is even more surprising given the ubiquity of time dependent data in many scientific disciplines. Data in fields such as climate science, neuroscience, political science, economics, and finance are frequently observed over time and/or space thereby exhibiting serial dependence. A specific example is in forecasting low frequency macroeconomic indicators such as GDP or inflation rate, where we can have a large number of macroeconomic and financial time series and their lags as possible covariates. Another example is identification of brain connectivity networks, where we have data from thousands of voxels collected over a relatively small number of time periods \citep{valdes2005}. These examples, amongst others, highlight the importance of developing screening methods for time dependent data. 

In creating a screening method for time series data, we aim to account for some of the unique features of time series data such as:

\begin{itemize}[noitemsep,nolistsep]
\item A prior belief that a certain number of lags of the response variable are to be in the model.
\item An ordered structure of the covariates, in which lower order lags of covariates are thought to be more informative than higher order lags.
\item The frequent occurrence of multivariate response models such linear or nonlinear VAR models.
\end{itemize}
Additionally, we aim to have a model free screening approach which can handle continuous, discrete or grouped time series. Using a model free approach makes our methods robust to model misspecification at the screening stage, and gives us full flexibility when considering a second stage procedure. The few works which have relaxed the assumption of independent observations include \cite{Cheng2014}, and \cite{XuB2014} which dealt with longitudinal data. However, the dependence structure of longitudinal data is too restrictive to cover the type of dependence present in most time series. To the best of our knowledge there have been only two works, \cite{Chenlu2017} and \cite{Yousuf2017}, dealing with the issue in a general stationary time series setting. The former work extended the nonparametric independence screening approach used for independent observations to the time series setting. However, the method does not utilize the serial dependence in the data, or account for the unique properties of time series data we outlined. The latter work \citep{Yousuf2017} extended the theory of SIS to heavy tailed and/or dependent data as well as proposing a GLS based screening method to correct for serial correlation. However, this work is limited to the linear model and the other unique qualities of time series data outlined above are ignored. Additionally both of these works are only applicable to models with a univariate response. 

In order to account for the unique characteristics of time series data mentioned above, and correct some of the limitations in previous works, we will introduce several distance correlation based screening procedures. Distance correlation (DC) was introduced by \cite{Szekeley2007}, for measuring dependence and testing independence between two random vectors. The consistency, and weak convergence of sample distance correlation has been established for stationary time series in \cite{Zhou12} and \cite{DavisWan16}. DC has a number of useful properties such as: 
\begin{itemize}[noitemsep,nolistsep]
\item The distance correlation of two random vectors equals to zero if and only if these two random vectors are independent.
\item Ability to handle discrete time series, as well as grouped predictors. 
\item An easy to compute partial distance correlation has also been developed, allowing us to control for the effects of a multivariate random vector \citep{Szekeley2014}. 
\end{itemize}
The first property allows us to develop a model free screening approach, which is robust to model misspecification. The second property is useful when dealing with linear or nonlinear VAR models for discrete or continuous data. The third property will allow us to account for the first two unique features of time series data mentioned previously. 

Compared to the previous works on screening using distance correlation based methods \citep{Lietal2012,liu2017}, our work differs in a number of ways. First, our work deals with the time series setting, where both the covariates and response are stationary time series, and can be heavy tailed. Second, our screening procedures are developed specifically in order to account for certain unique features in time series data mentioned previously. Lastly, we choose to rely on partial DC, instead of conditional DC, when controlling for confounding variables. Partial DC is a DC based procedure which can be easily computed using pairwise distance correlations, whereas the computation of conditional DC is more involved and involves the choice of a bandwidth parameter, which can be difficult to choose. 

Broadly speaking, we will be dealing with two types of models: univariate response models, some examples of which include linear or nonlinear autoregressive models with exogenous predictors (NARX), and multivariate response models such as linear or nonlinear VAR models. In both settings, we rely on partial distance correlation to build our screening procedures. Partial distance correlation produces a rich family of screening methods by taking different choices for the conditioning vector. In many applications, it is usually the case that researchers have prior knowledge that a certain subset of predictors is relevant to the response. Utilizing this prior knowledge usually enhances the screening procedure, as shown in the case of generalized linear models in \cite{Fan2016}. Therefore our procedure can be viewed as a model free adaption of this principle to the time series setting. We discuss approaches for choosing the conditioning vector of each predictor, and we usually assume at least a few lags of the response variable are part of the  conditioning vector of each predictor. We also discuss ways in which we can leverage the ordered structure of our lagged covariates to add additional variables to our conditioning vectors. 

To motivate the multivariate response setting, consider a linear VAR(1) model: $\bm{x}_t= B_1\bm{x}_{t-1} + \bm{\eta}_t$, where $\bm{x}_t$ is a $p$-variate random vector. The number of parameters to estimate in this model is $p^2$, which can quickly become computationally burdensome even for screening procedures.  In many cases however, there exists a certain group structure amongst the predictors, which is known to researchers in advance, along with a sparse conditional dependency structure between these groups \citep{basu15}. For example, in macroeconomics or finance, different sectors of the economy can be grouped into separate clusters. Using this group structure, we can apply the partial distance correlation to screen relationships at the group level, thereby quickly reducing the number of variables for a second stage procedure.

The rest of the paper is organized as follows. Section \ref{sec:section 2} reviews the functional dependence measure and $\beta$ mixing coefficients, as well as comparisons between the two frameworks. We also discuss the assumptions placed on structure of the covariate and response processes.
Section \ref{sec:section 3} introduces our screening procedures with their sure screening properties for models with a univariate response. Section \ref{section4} presents screening algorithms for multivariate response models. Section \ref{section5} covers simulation results, and a real data application is presented in Section \ref{section6}. The concluding remarks are in Section \ref{section7}. Lastly, the proofs for all theorems, along with additional simulations and data analysis results for sections \ref{section5} and \ref{section6} are placed in the supplementary material.

\section{Dependence Measures}\label{sec:section 2}

In order to establish asymptotic properties, we rely on two widely used dependence measures, the functional dependence measure and $\beta$-mixing coefficients. We first start with an overview of the functional dependence measure framework, before proceeding to $\beta$-mixing processes. For univariate processes, $(Y_i \in \mathcal{R})_{i \in \mathbb{Z}}$, we assume $Y_i$ is a causal, strictly stationary, ergodic process with the following form:
\begin{equation}Y_i=g\left(\ldots,e_{i-1},e_i\right)\label{nonlinear},\end{equation} 
where $g(\cdot)$ is a real valued measurable function, and $e_i$ are iid random variables. 
And for multivariate processes, such as the covariate process $(\bm{x}_i \in \mathcal{R}^{p_n})_{i \in \mathbb{Z}}$, we assume the following representation:
\begin{align}\bm{x_i}=\bm{h}\left(\ldots,\bm{\eta}_{i-1},\bm{\eta}_i\right)\label{nonlinear2}.\end{align} 
Where $\bm{\eta}_i, i\in \mathbb{Z}$, are iid random vectors, $\bm{h}(\cdot)=(h_1(\cdot)\ldots,h_{p_n}(\cdot))$, $\bm{x_i}=(X_{i1},\ldots,X_{ip_n})$, and $X_{ij}=h_j(\ldots,\bm{\eta}_{i-1},\bm{\eta}_{i})$.

Processes having these representations are sometimes known as Bernoulli shift processes \citep{Wu2009SPA}, and include a  wide range of stochastic processes such as linear processes with their nonlinear transforms, Volterra processes, Markov chain models, nonlinear autoregressive models such as threshold auto-regressive (TAR), bilinear, GARCH models, among others  \citep{Wu2011, Wu2005}. These representations allow us to quantify dependence using a functional dependence measure introduced in \cite{Wu2005}. The functional dependence measure for a univariate process and multivariate processes is defined respectively as:
\begin{eqnarray}\delta_{q}(Y_i)=||Y_i-g\left(\mathcal{F}_i^*\right)||_q = (E|Y_i-g\left(\mathcal{F}_i^*\right)|^q)^{1/q},\nonumber\\
\delta_{q}(X_{ij})=||X_{ij}-h_{j}\left(\mathcal{H}_i^*\right)||_q = (E|X_{ij}-h_{j}\left(\mathcal{H}_i^*\right)|^q)^{1/q},\end{eqnarray}
where $\mathcal{F}_i^*=\left(\ldots,e_{-1},e_0^*,e_1,\ldots,e_i\right)$ with $e_0^*,e_j, j \in \mathbb{Z} $ being iid. 
And for the multivariate case, $\mathcal{H}_{i}^{*}=(\ldots,\bm{\eta}_{-1},\bm{\eta}^{*}_{0},\bm{\eta}_{1},\ldots,\bm{\eta}_{i})$ with $\bm{\eta}_0^*,\bm{\eta}_j, j \in \mathbb{Z} $ being iid. Since we are replacing $e_0$ by $e_0^*$, we can think of this as measuring the dependency of $y_i$ on $e_0$, since we are keeping all other inputs the same. We assume the cumulative functional dependence measures are finite:
\begin{equation}\Delta_{0,q}(\bm{y})=\sum_{i=0}^{\infty}\delta_{q}(Y_i)<\infty,\textrm{ and }
\Phi_{m,q}(\bm{x})=\max_{j \leq p_n}\sum_{i=m}^{\infty}\delta_{q}(X_{ij}) < \infty \label{qstrongstable} .\end{equation}
This short range dependence condition implies, by the proof of theorem 1 in \cite{Wu2009}, the auto-covariances are absolutely summable.

There are many equivalent definitions given for $\beta$-mixing, and we use the one provided by \cite{Doukhan94}: 
\begin{defn} Given a stationary multivariate process, $(\bm{x_i})_{i \in \mathcal{Z}}$, for each positive integer $a$, the coefficient of absolute regularity or $\beta$-mixing coefficient, $\beta(a)$, is:
$$ \beta(a)=||\mathcal{P}_{-\infty:0} \times \mathcal{P}_{a:\infty}-\mathcal{P}_{-\infty:0,a:\infty}||_{TV}. $$
Where $||\cdot||_{TV}$ is the total variation norm, and $\mathcal{P}_{-\infty:0,a:\infty}$ is the joint distribution of the blocks $(\bm{x}_{-\infty:0}, \bm{x}_{a,\infty})$. A stochastic process is said to be $\beta$-mixing if $\beta(a) \rightarrow 0.$

\end{defn}

We note that compared to functional dependence measures, $\beta$-mixing coefficients can be defined for any stochastic processes, and are not limited to Bernoulli shift processes. On other hand, functional dependence measures are easier to interpret and compute since they are related to the data generating mechanism of the underlying process. In many cases using the functional dependence measure also requires less stringent assumptions (see \cite{WuandWu2016}, \cite{Yousuf2017} for details). Although there is no direct relationship between these two dependence frameworks, fortunately there are a large number of commonly used time series processes which are $\beta$-mixing and satisfy (\ref{qstrongstable}). For example, under appropriate conditions, linear processes, ARMA, GARCH, ARMA-ARCH, threshold autoregressive, Markov chain models, amongst others, can be shown to be $\beta$-mixing (see \cite{Pham85}, \cite{Carrasco02}, \cite{An96}, \cite{lu1998} for details). 
%

\section{Partial Distance Correlation Screening for Univariate Response Models}\label{sec:section 3}
\subsection{Preliminaries}
We start with a brief overview of the distance covariance, distance correlation, and partial distance correlation measures. 
\begin{defn} For any random vectors $\bm{u} \in \mathcal{R}^q,\bm{v} \in \mathcal{R}^p$, let $\phi_{\bm{u}}(\bm{t}), \phi_{\bm{v}}(\bm{s})$ be the characteristic function of $\bm{u}$ and $\bm{v}$ respectively. The distance covariance between $\bm{u}$ and $\bm{v}$ is defined as \cite{Szekeley2007}:
$$ dcov^2(\bm{u},\bm{v})= \int_{\mathcal{R}^{p+q}} |\phi_{\bm{u,v}}(\bm{t,s})-\phi_{\bm{u}}(\bm{t})\phi_{\bm{v}}(\bm{s})|^2 \omega^{-1}(\bm{t,s}) d\bm{t} d\bm{s}, $$
where the weight function $\omega(\bm{t,s})=c_pc_q|\bm{t}|_p^{1+p}|\bm{s}|_q^{1+q}$, where $c_p=\frac{\pi^{(1+p)/2}}{\Gamma((1+p)/2)}$. Throughout this article $|\bm{a}|_p$ stands for the Euclidean norm of $ \bm{a} \in \mathcal{R}^p$.
\end{defn}
Given this choice of weight function, by \cite{Szekeley2007}, we have a simpler formula for the distance covariance. Let $(\bm{u,v}),(\bm{u',v'}),(\bm{u'',v''})$ be iid, each with joint distribution $(\bm{u,v})$, and let:
$$S_1=E(|\bm{u}-\bm{u}'|_p|\bm{v}-\bm{v}'|_q),\textrm{ }S_2=E(|\bm{u}-\bm{u}'|_p)E(|\bm{v}-\bm{v}'|_q),\textrm{ } S_3=E(|\bm{u}-\bm{u}'|_p)E(|\bm{v}-\bm{v}''|_q).$$
Then, by remark 3 in \cite{Szekeley2007}, $dcov^2(\bm{u},\bm{v})= S_1 + S_2 - 2S_3.$ We can now estimate this quantity using moment based methods. Suppose we observe $(\bm{u}_i,\bm{v}_i)_{i=1,\ldots,n}$, the sample estimates for the distance covariance and distance correlation are:
\begin{align*} &\widehat{dcov^2}(\bm{u},\bm{v})=\hat{S}_1 + \hat{S}_2 - 2\hat{S}_3 \textrm{, and }\widehat{dcor}(\bm{u},\bm{v})=\frac{\widehat{dcov}(\bm{u},\bm{v})}{\sqrt{\widehat{dcov}(\bm{u},\bm{u})\widehat{dcov}(\bm{v},\bm{v})}}, \\
\textrm{where }&\hat{S}_1=n^{-2}\sum_{i,j=1}^{n}|\bm{u}_i-\bm{u}_j|_p|\bm{v}_i-\bm{v}_j|_q,\quad\hat{S}_2=n^{-2}\sum_{i,j=1}^{n}|\bm{u}_i-\bm{u}_j|_pn^{-2}\sum_{i,j=1}^{n}|\bm{v}_i-\bm{v}_j|_q,\nonumber \\
&\hat{S}_3=n^{-3}\sum_{i,j,l=1}^{n}|\bm{u}_i-\bm{u}_j|_p |\bm{v}_i-\bm{v}_l|_q .\end{align*}

Partial distance correlation (PDC) was introduced in \cite{Szekeley2014}, as a means of measuring nonlinear dependence between two random vectors $\bm{u}$ and $\bm{v}$ while controlling for the effects of a third random vector $\bm{\mathcal{Z}}$. We refer to the vector $\bm{\mathcal{Z}}$ as the conditioning vector. Additionally, \cite{Szekeley2014} showed that the PDC can be evaluated using pairwise distance correlations. Specifically, the PDC between $\bm{u}$ and $\bm{v}$, controlling for $\bm{\mathcal{Z}}$, is defined as:
$$pdcor(\bm{u,v;\mathcal{Z}})=\frac{dcor^2(\bm{u,v})-dcor^2(\bm{u,\mathcal{Z}})dcor^2(\bm{v,\mathcal{Z}})}{\sqrt{1-dcor^4(\bm{u,\mathcal{Z}})}\sqrt{1-dcor^4(\bm{v,\mathcal{Z}})}}, $$
if $dcor(\bm{u,\mathcal{Z}}),dcor(\bm{v,\mathcal{Z}}) \neq 1$, otherwise $pdcor(\bm{u,v;\mathcal{Z}})=0$. For more details and an interpretation of PDC, one can consult \cite{Szekeley2014}.
\subsection{Screening Algorithm I: PDC-SIS}
We first review some basic ingredients of screening procedures. Let $\bm{y} =\left(Y_1,\ldots,Y_n\right)^T$ be our response time series, and let $\bm{x_{t-1}}=(X_{t-1,1},\ldots,X_{t-1,m_n})$ denote the $m_n$ predictor series at time $t-1$. Given that lags of these predictor series are possible covariates, we let $\bm{z_{t-1}}=(\bm{x_{t-1},x_{t-2},\ldots,x_{t-h}})=(Z_{t-1,1},\ldots,Z_{t-1,p_n})$ denote the length $p_n$ vector of covariates, where $p_n=m_n\times h$. Now we denote our set of active covariates as:
$$\mathcal{M}_*=\left\{j \leq p_n: F(Y_t|Y_{t-1},\ldots,Y_{t-h},\bm{z_{t-1}}) \textrm{ functionally depends on } Z_{t-1,j} \right\},$$
where $F(Y_t|\cdot)$ is the conditional cumulative distribution function of $Y_t$. The value $h$ represents the maximum lag order we are considering for our response and predictor series. This value can be decided beforehand by the user, or can be selected using a data driven method. Variable selection methods aim to recover $\mathcal{M}_*$ exactly, which can be a very difficult goal both computationally and theoretically, especially when $p_n \gg n$. In contrast, variable screening methods have a less ambitious goal, and aim to find a set $S$ such that $P(\mathcal{M}_* \subset S) \rightarrow 1$ as $n \rightarrow \infty$. Ideally we would also hope that $|S| \ll p_n$, thereby significantly reducing the dimension of the feature space for a second stage method. 

When developing screening algorithms for time series data, we would like to account for some of its unique properties as mentioned in the introduction. For models with a  response, these would be: 
\begin{itemize}[noitemsep]
\item A prior belief that a certain number of lags of the response variable are to be in the model.
\item An ordered structuring of the covariates, in which lower order lags of covariates are thought to be more informative than higher order lags.
\end{itemize}
The first property can be easily accounted for using partial distance correlation, while there are many different ways to account for the second property. In this section we present two partial distance correlation based screening algorithms, which attempt to account for the ordered structure of our covariates. In our first algorithm, PDC-SIS, we define the conditioning vector for the $l^{th}$ lag of predictor series $k$ as:
$$\mathcal{S}_{k,l}=(Y_{t-1},\ldots,Y_{t-h},X_{t-1,k},\ldots,X_{t-l+1,k}), $$
where $1 \leq l\leq h$. Since we are assuming \textit{a priori} that a certain number of lags of $Y_t$ are to be included in the model, $\left\{Y_{t-1},\ldots,Y_{t-h}\right\}$ is part of the  conditioning vector for all possible covariates. Our conditioning vector also includes all lower order lags for each lagged covariate we are considering. By including the lower order lags in the  conditioning vector, our method tries to shrink towards sub-models with lower order lags. To illustrate this, consider the case where $Y_t$ is strongly dependent on $X_{t-1,j}$ even while controlling for the effects of $Y_{t-1},\ldots, Y_{t-h}$. Under this scenario, if $X_{t-1,j}$ has strong serial dependence, higher order lags of $X_{t-1,j}$ can be mistakenly  selected by our screening procedure even if they are not in our active set of covariates. 

For convenience, let $\bm{C}=\left\{\mathcal{S}_{1,1},\ldots,\mathcal{S}_{m_n,1},\mathcal{S}_{1,2},\ldots,\mathcal{S}_{m_n,h}\right\}$ denote our set of  conditioning vectors; where $C_{k+(l-1)*m_n}=\mathcal{S}_{k,l}$ is the conditioning vector for covariate $Z_{t-1,(l-1)*m_n + k}$. Our screened sub-model is:
$$\hat{\mathcal{M}}_{\gamma_n}=\left\{j \in \left\{1,\ldots,p_n\right\}:|\widehat{pdcor}(Y_t,Z_{t-1,j};C_j)| \geq \gamma_n \right\}.$$
To establish sure screening properties, we introduce the following conditions.
\begin{condition}\label{ConditionA}
Assume $|pdcor(Y_t,Z_{t-1,k};C_k)| \geq c_1 n^{-\kappa}$ for $k \in M_*$ and $\kappa \in (0,1/2)$.
\end{condition}
\begin{condition}\label{ConditionB}
Assume the response and the covariate processes have representations (\ref{nonlinear}) and (\ref{nonlinear2}), respectively. Additionally, we assume the following decay rates $\Phi_{m,r}(\bm{x}) = O(m^{-\alpha_z}), \Delta_{m,q}(\bm{y}) =O(m^{-\alpha_{y}})$, for some $\alpha_z, \alpha_{y} > 0$, $q > 2, r > 4 $ and $\tau=\frac{qr}{q+r} > 2$. 
\end{condition}
\begin{condition}\label{ConditionC} 
Assume the response and the covariate processes have representations (\ref{nonlinear}) and (\ref{nonlinear2}) respectively. Additionally assume \\$\upsilon_z=\sup_{q \geq 2} q^{-\tilde{\alpha}_z}\Phi_{0,q}(\bm{x}) < \infty$ and $\upsilon_{y}=\sup_{q \geq 2} q^{-\tilde{\alpha}_{y}} \Delta_{0,q}(\bm{y})< \infty$, for some $\tilde{\alpha}_z,\tilde{\alpha}_{y} \geq 0.$
\end{condition}
\begin{condition}\label{ConditionD}
Assume the process $\left\{(Y_t,\bm{x}_t)\right\}$ is $\beta$-mixing, with mixing rate $\beta_{xy}(a)=O(\exp(-a^{\lambda_1}))$, for some $\lambda_{1} >0$.
\end{condition}
Condition \ref{ConditionA} is a standard population level assumption which allows covariates in the active set to be detected by our screening procedure. Condition \ref{ConditionB} is similar to the one used in \cite{Yousuf2017} and \cite{WuandWu2016}, and assumes both the response and covariate processes are causal Bernoulli shift processes. Additionally it presents the dependence and moment conditions on these processes, where higher values of $\alpha_x,\alpha_{\epsilon}$ indicate weaker temporal dependence. Examples of response processes which satisfy condition \ref{ConditionB} 
include stationary, causal, finite order ARMA, GARCH, ARMA-GARCH, bilinear, and threshold autoregressive processes, all of which have exponentially decaying functional dependence measures (see \cite{Wu2011} for details). For the covariate process, assume $\bm{x}_i$ is a vector linear process: $\bm{x}_i=\sum_{l=0}^{\infty} A_l\bm{\eta}_{i-l}$. where $\{A_{l}\}$ are $m_n \times m_n$ coefficient matrices and $\{\bm{\eta}_i=(\eta_{i1},\ldots,\eta_{im_n})\}$ are iid random vectors with $cov(\bm{\eta_i})=\Sigma_{\eta}$. For simplicity, assume $\{\eta_{i,j}, j=1,\ldots,m_n\}$ are identically distributed, then 
$\delta_{q}(X_{ij})=||A_{i,j}\bm{\eta}_0-A_{i,j}\bm{\eta}_0^{*}||_q \leq 2|A_{i,j}|||\eta_{0,1}||_q$, 
where $A_{i,j}$ is the $j^{th}$ column of $A_i$. If $||A_i||_{\infty}=O(i^{-\beta})$ for $\beta>1$, then $\Phi_{m,q}(\bm{x})=O(m^{-\beta+1})$. Other examples include stable VAR processes, and multivariate ARCH processes which have exponentially decaying cumulative functional dependence measures \citep{WuandWu2016,Yousuf2017}. Condition \ref{ConditionC} strengthens the moment requirements of condition \ref{ConditionB}, and requires that all moments of the covariate and response processes are finite. To illustrate the role of the constants $\tilde{\alpha}_{z}$ and $\tilde{\alpha}_{y}$, consider the example where $y_i$ is a linear process: $y_i=\sum_{j=0}^{\infty} f_j e_{i-j}$ with $e_i$ iid and $\sum_{l=0}^{\infty}|f_l|<\infty$, then $\Delta_{0,q}(\bm{y})= ||e_{0}-e_{0}^{*}||_{q}\sum_{l=0}^{\infty}|f_l|$. If we assume $e_0$ is sub-Gaussian, then $\tilde{\alpha}_{y}=1/2$, since $||e_{0}||_{q} = O(\sqrt{q})$. Similarly, if $e_i$ is sub-exponential, we have $\tilde{\alpha}_{y}=1$.

To understand the inclusion of condition \ref{ConditionD}, consider the $U$-statistic:
$$U_r(S_{t_1},\ldots,S_{t_r})=\binom{n}{r}\sum_{t_1 \leq t_2 \leq \ldots \leq t_r \leq n} h(S_{t_1},\ldots,S_{t_r}),$$ which aims to estimate $\theta(h)=\int h(S_{t_1},\ldots,S_{t_r}) d\mathcal{P}(S_{1})\ldots d\mathcal{P}(S_r)$. When $S_1,\ldots, S_n$ are iid, the $U$-statistic is an unbiased estimator of $\theta(h)$, however for $r>1$ the $U$-statistic is no longer unbiased if $S_t$ is serially dependent. Since our sample distance correlation estimate can be written as a sum of $U$-statistics \citep{Lietal2012}, condition \ref{ConditionD} is needed to control the rate at which the above bias vanishes as $n \rightarrow \infty$. Conditions \ref{ConditionB} and \ref{ConditionD} are frequently used when dealing with time series data \citep{Wu2009,Wu2012,DavisWan16}.

Throughout this paper, let $\alpha = \min(\alpha_x,\alpha_{y})$, and $\varrho =1$, if $\alpha_z > 1/2 - 2/r$, otherwise $\varrho= r/4 - \alpha_z r/2$. Let $\iota = 1$ if $\alpha > 1/2-1/\tau$, otherwise $\iota = \tau/2-\tau\alpha$, and let $\zeta =1$, if $\alpha_y> 1/2 - 2/q$, otherwise $\zeta= q/4 - \alpha_y q/2$. Additionally, let $K_{y,q}=\sup_{m \geq 0} (m+1)^{\alpha_{y}}\Delta_{m,q}(\bm{y})$, and $K_{z,r}=\sup_{m \geq 0} (m+1)^{\alpha_z} \Phi_{r}(\bm{x})$. Given condition \ref{ConditionC}, it follows that $K_{\epsilon,q}, K_{z,r} < \infty$. Let $t_n=\max_j dim(C_j)$, be the maximum dimension of the conditional vectors. We define $\tilde{\psi}=\frac{2}{1+2\tilde{\alpha}_z+2\tilde{\alpha}_y}, \tilde{\varphi} = \frac{2}{1+4\tilde{\alpha}_z}$, $\tilde{\alpha} = \frac{2}{1+4\tilde{\alpha}_y}$. Lastly, for ease of presentation, let $\hat{\bm{\omega}}=(\hat{\omega}_1,\ldots,\hat{\omega}_{p_n})$, $\bm{\omega}=(\omega_1,\ldots,\omega_{p_n})$, where $\omega_k=pdcor(Y_t,Z_{t-1,k};C_k)$, $\hat{\omega}_k=\widehat{pdcor}(Y_t,Z_{t-1,k};C_k)$. In addition,  let 
\begin{align*}
a_n&= n^2\left[\exp\left(-\frac{n^{1/2-\kappa}}{t_n\upsilon_y^2}\right)^{\tilde{\alpha}}+ \exp\left(-\frac{n^{1/2-\kappa}}{t_n\upsilon_z\upsilon_{y}}\right)^{\tilde{\psi}}+ \exp\left(-\frac{n^{1/2-\kappa}}{t_n\upsilon_z^2}\right)^{\tilde{\varphi}}\right], \nonumber\\
b_n&=n^2\Bigg[\frac{t_n^{r/2}n^{\zeta} K_{y,r}^{r}}{n^{r/2-r\kappa/2}} +\frac{t_n^{r/2}n^{\iota} K_{z,r}^{r/2}K_{y,r}^{r/2}}{n^{r/2-r/2\kappa}}+\frac{t_n^{r/2}n^{\varrho} K_{z,r}^{r}}{n^{r/2-r\kappa/2}}\\
&\quad+\exp\left(-\frac{n^{1-2\kappa}}{t_n^2K_{z,r}^4}\right)+ \exp\left(-\frac{n^{1-2\kappa}}{t_n^2K_{z,r}^2K_{y,r}^2}\right)+ \exp\left(-\frac{n^{1-2\kappa}}{t_n^2K_{y,r}^4}\right)\Bigg],\\
c_n&= \frac{t_n^{r/2}K_{y,r}^{r}}{n^{r/4-r\kappa/2}}+\frac{t_n^{r/2}K_{z,r}^{r/2}K_{y,r}^{r/2}}{n^{r/4-r/2\kappa}}+\frac{t_n^{r/2}K_{z,r}^{r}}{n^{r/4-r\kappa/2}}.
\end{align*}

For simplicity and convenience of presentation, we assume $q=r$, and one can consult the proof for the general case. The following theorem presents the sure screening properties of PDC-SIS for both the heavy tailed and light tailed settings. 

\begin{thm}{}   
\begin{enumerate}[noitemsep]
\item Suppose conditions \ref{ConditionA}, \ref{ConditionC}, and \ref{ConditionD} hold. For any $c_2 > 0$, we have:
$$P(\max_{j \leq p_n}|\hat{\omega}_k -\omega_k|>c_2 n^{-\kappa}) \leq O(p_na_n).$$

\item Suppose conditions \ref{ConditionA}, \ref{ConditionC}, and \ref{ConditionD} hold. For $\gamma_n = c_3 n^{-\kappa}$ with $c_3 \leq c_1/2$, we have:
$$P\left(\mathcal{M}_*\subset \hat{\mathcal{M}}_{\gamma_n} \right)\geq 1-O(s_na_n).$$
\item Suppose conditions \ref{ConditionA}, \ref{ConditionB}, and \ref{ConditionD} hold. For any $c_2 > 0$, we have:
$$\mbox{ if $r<12$,} \qquad P(\max_{j \leq p_n}|\hat{\omega}_j -\omega_j|>c_2 n^{-\kappa}) \leq O(p_nc_n);$$
$$\mbox{ if $r\geq12$,} \qquad P(\max_{j \leq p_n}|\hat{\omega}_k -\omega_k|>c_2 n^{-\kappa}) \leq O(p_nb_n)
.$$
\item Suppose conditions \ref{ConditionA}, \ref{ConditionB}, and \ref{ConditionD} hold. For $\gamma_n = c_3 n^{-\kappa}$ with $c_3 \leq c_1/2$, we have:
$$\mbox{ if $r<12$,} \qquad P\left(\mathcal{M}_*\subset \hat{\mathcal{M}}_{\gamma_n} \right)\geq 1-O(s_nc_n)
;$$
$$\mbox{ if $r\geq12$,} \qquad P\left(\mathcal{M}_*\subset \hat{\mathcal{M}}_{\gamma_n} \right)\geq 1-O(s_nb_n).$$
\end{enumerate}
\label{theorem1} 
\end{thm}

From the above theorem, we observe that the range of $p_n$ depends on the temporal dependence in both the covariate and the response processes, the strength of the signal ($\kappa$), and the moment conditions. We also have two cases for finite polynomial moments, one for $r<12$ and one for $r\geq 12$. This is due to our proof technique which relies on both Nagaev and Rosenthal type inequalities. For the case of low moments, we obtain a better bound using a Rosenthal type inequality combined with the Markov inequality, whereas for higher moments Nagaev type inequalities lead to a better bound; more details can be found in the proof which is provided in the supplementary file. 

For example, if we assume only finite polynomial moments with $r=q$ and $r<12$, then $p_n=o(n^{r/4-r\kappa/2})$. If we assume $\alpha \geq 1/2 - 2/r$ and $r>12$,  $p_n=o(n^{r/2-r\kappa/2-3})$. The constants $K_{z,r}$ and $K_{y,q}$, which are related to the cumulative functional dependence measures, represent the effect of temporal dependence on our bounds when $\alpha \geq 1/2 - 2/r$. However, when using Nagaev type inequalities, there is an additional effect in the case of stronger dependence in the response or covariate process (i.e. $\alpha < 1/2 - 2/r$). For instance, if $\alpha_x=\alpha_{\epsilon}$ and $q=r$, the range for $p_n$ is reduced by a factor of $n^{r/4-\alpha r/2}$ in the case of stronger dependence. For the case of exponentially decaying tails however, there is no level shift in the decay rate of our bounds due to the dependence of the response or covariate processes. We observe that if the response and covariates are sub-Gaussian, $p_n=o(n^{\frac{1-2\kappa}{3}})$, and if they are sub-exponential, $p_n=o(n^{\frac{1-2\kappa}{5}})$. 

By choosing an empty conditional set for all the variables, our procedure reduces to the distance correlation screening (DC-SIS) introduced in \cite{Lietal2012} for the iid setting. Assuming sub-Gaussian response and covariates, \cite{Lietal2012} obtained $p_n=o(n^{\frac{1-2\kappa}{3}})$ for DC-SIS, which matches our rate. In the iid setting with finite polynomial moments, we can use the truncation method in their proof and combined with the Markov inequality to obtain $p_n=o(n^{r/4-r\kappa/2-1})$. Our results, which rely on a different proof strategy than the truncation method, provide a better bound even in this setting.

\subsection{Screening Algorithm II: PDC-SIS+}

As we have seen, the time ordering of the covariates allows us some additional flexibility in selecting the conditioning vector compared to iid setting. Our previous algorithm attempted to utilize the time series structure of our data by conditioning on previous lags of the covariate. However, rather than simply conditioning only on the previous lags of a covariate, we can condition on additional information available from previous lags of other covariates as well. One way to attempt this, and to potentially improve our algorithm, is to identify strong conditional signals at each lag level and add them to the conditioning vector for all higher order lag levels. By utilizing this conditioning scheme we can pick up on hidden significant variables in more distant lags, and also shrink toward models with lower order lags by controlling for false positives resulting from high autocorrelation, and cross-correlation.

We now give a formal description of PDC-SIS+. The conditioning vector for the first lag level of predictor series $k$ is:
$\mathcal{S}_{k,1}=(Y_{t-1},\ldots,Y_{t-h}), $ 
which  coincides with the conditioning vector for the first lag level of PDC-SIS. Using the representation $\bm{z}_{t-1}=(\bm{x_{t-1}},\ldots,\bm{x}_{t-h})$, we denote the strong conditional signal set for the first lag level as:
$$\hat{\mathcal{U}}_{1}^{\lambda_n}=\left\{j \in \left\{1,\ldots,m_n\right\}:|\widehat{pdcor}(Y_t,Z_{t-1,j};\mathcal{S}_{j,1})| \geq \lambda_n+c_1n^{-\kappa} \right\}.$$
We then use this information to form our next conditioning vector:
$$ \hat{\mathcal{S}}_{k,2}=\left(Y_{t-1},\ldots,Y_{t-h},X_{t-1,k},\bm{z}_{t-1,\hat{\mathcal{U}}_{t-1}^{\lambda_n}}\right), $$
where $\bm{z}_{t-1, \hat{\mathcal{U}}_{1}^{\lambda_n}}$ is a sub-vector of $\bm{z}_{t-1}$ which is formed by extracting the indices contained in $\hat{\mathcal{U}}_{1}^{\lambda_n}$.
We note that any duplicates which result from overlap between ${X_{t-1,k}}$ and ${\bm{z}_{t-1,\hat{\mathcal{U}}_{1}^{\lambda_n}}}$ are deleted. For convenience, we define $\hat{C}=(\hat{\mathcal{S}}_{1,1},\ldots,\hat{\mathcal{S}}_{m_n,1},\hat{\mathcal{S}}_{1,2},\ldots,\hat{\mathcal{S}}_{m_n,h})$ as our vector of estimated conditional sets. We then use $(\hat{\mathcal{S}}_{k,2})_{k \leq m_n}$ to compute the strong conditional signal set for the $2^{nd}$ lag level:
\begin{equation*} \hat{\mathcal{U}}_{2}^{\lambda_n}=\left\{j \in \left\{m_n+1,\ldots,2m_n\right\}:|\widehat{pdcor}(Y_t,Z_{t-1,j};\hat{C}_{j})| \geq \lambda_n +c_1n^{-\kappa}\right\}.\end{equation*} 
Repeating this procedure we obtain:
\begin{equation*} \hat{\mathcal{S}}_{k,l}=\left(Y_{t-1},\ldots,Y_{t-h},X_{t-1,k},\ldots,X_{t-l+1,k},\bm{z}_{t-1,\hat{\mathcal{U}}_{1}^{\lambda_n}},\ldots,\bm{z}_{t-1,\hat{\mathcal{U}}_{l-1}^{\lambda_n}}\right). \end{equation*}
We can also vary the threshold $\lambda_n$ for each lag level; for simplicity we leave it the same for each of our levels here. Our sub-model obtained from this procedure is: 
\begin{equation*} \widetilde{\mathcal{M}}_{\gamma_n}=\left\{j \in \left\{1,\ldots,p_n\right\}:|\widehat{pdcor}(Y_t,Z_{t-1,j};\hat{C}_{j})| \geq \gamma_n \right\}.\end{equation*} 
The asymptotic properties of this procedure are similar to PDC-SIS, and we present them in the supplementary material.
To show the asymptotic properties associated with this algorithm, we denote
 $$\mathcal{S}_{k,l}=\left(Y_{t-1},\ldots,Y_{t-h},X_{t-1,k},\ldots,X_{t-l+1,k},\bm{z}_{t-1,\mathcal{U}_{1}^{\lambda_n}},\ldots,\bm{z}_{t-1,\mathcal{U}_{l-1}^{\lambda_n}} \right),$$ as the population level counterpart to $\hat{\mathcal{S}}_{k,l}$. In addition, let $C=\{\mathcal{S}_{1,1},\ldots,\mathcal{S}_{m_n,1},\mathcal{S}_{1,2},$ $\ldots,\mathcal{S}_{m_n,h}\},$
 and 
\begin{eqnarray} \mathcal{U}_{l-1}^{\lambda_n}=&\left\{(l-1)m_n+1\le j\le lm_n:|pdcor(Y_t,Z_{t-1,j};C_{j})| \geq \lambda_n+\frac{c_1}{2}n^{-\kappa} \right\},\nonumber
\end{eqnarray}
represent the population level strong conditional signal set and the population level set of conditioning vectors, respectively. One of the difficulties in proving uniform convergence of our estimated partial distance correlations in this algorithm is the presence of an estimated conditioning set $\hat{{C}}$. This issue becomes compounded as we estimate the conditioning vector for higher lag levels, since these rely on estimates of the conditioning vectors for lower ones. To overcome this, we first denote the collection of strong signals from lag 1  to $h-1$ as: $\mathcal{U}^{\lambda_n}=\left\{\mathcal{U}_{1}^{\lambda_n},\ldots,\mathcal{U}_{h-1}^{\lambda_n}\right\}$. We will assume the following condition:
\begin{condition}\label{ConditionE} For any $j \in \left\{1,\ldots,(h-1)*m_n\right\} \setminus \mathcal{U}^{\lambda_n}$, assume \\ $|pdcor(Y_t,Z_{t-1,j};C_{j})| \leq \lambda_n$, where ${\lambda_n}{n^{\kappa}} \rightarrow \infty$.
\end{condition}
Condition \ref{ConditionE} assumes the variables in the strong conditional signal set, $\mathcal{U}^{\lambda_n}$, are easily identifiable from the rest of the covariates. This separation in the signal strength will allow us to ensure with high probability that our estimated conditional sets match their population level counterparts. The assumption ${\lambda_n}{n^{\kappa}} \rightarrow \infty$, is introduced to ensure $d_n=|\widetilde{\mathcal{M}}_{\gamma_n}|\gg |\mathcal{U}^{\lambda_n}|$. Although the hope is that $\mathcal{U}^{\lambda_n} \subset \mathcal{M}_* $, this is not required to prove sure screening properties of our algorithm. Additionally, as seen in \cite{Fan2016} for the case of generalized linear models, conditioning on irrelevant variables could also enhance the power of a screening procedure. We will discuss how to choose the threshold $\Gamma_n=\lambda_n+c_1n^{-\kappa}$ for $\hat{\mathcal{U}}^{\lambda_n}$ in section \ref{Threshold}. In practice we would prefer not to condition on too many variables, therefore the threshold for adding a variable to $\mathcal{U}^{\lambda_n}$ would be high. 

The sure screening properties for PDC-SIS+ are similar to PDC-SIS, but for the sake of completeness, we state the theorem in full.

\begin{thm}{}   
\begin{enumerate}

\item Suppose conditions \ref{ConditionA}, \ref{ConditionC}, \ref{ConditionD}, and \ref{ConditionE} hold. For $\gamma_n = c_3 n^{-\kappa}$ with $c_3 \leq c_1/2$, we have
$$P\left(\mathcal{M}_*\subset \widetilde{\mathcal{M}}_{\gamma_n} \right)\geq 1-O(s_na_n).
$$
\item Suppose conditions \ref{ConditionA}, \ref{ConditionB}, \ref{ConditionD}, and \ref{ConditionE} hold. For $\gamma_n = c_3 n^{-\kappa}$ with $c_3 \leq c_1/2$, we have 
$$\mbox{ if $r<12$,} \qquad P\left(\mathcal{M}_*\subset \widetilde{\mathcal{M}}_{\gamma_n} \right)\geq 1-O(s_nc_n);
$$
$$\mbox{ if $r\ge12$,} \qquad P\left(\mathcal{M}_*\subset \widetilde{\mathcal{M}}_{\gamma_n} \right)\geq 1-O(s_nb_n). 
$$
\end{enumerate}
\label{theorem2} 
\end{thm}
Now, we have presented two classes of PDC screening methods. In the first class of methods, the conditional set of each covariate is known as \textit{a priori}, while in the second class the conditional set is estimated from the data. We can easily modify our algorithms for both procedures depending on the situation; for example we can screen groups of lags at a time for certain covariates in PDC-SIS. Additionally, for either procedure we can condition on a small number of lags of $Y_t$, and leave the higher order lags of $Y_t$ as possible covariates in our screening procedure.
\subsection{Threshold Selection: PDC-SIS+}
\label{Threshold}
In order to implement PDC-SIS+, we need to select a threshold parameter $\Gamma_n$. For simplicity we will only use a single threshold for all lag levels, and it is selected as follows: we first generate $1000$ independent AR(1) variables, $\xi_{t,1},\ldots,\xi_{t,1000}$ where $\xi_{t,i}=\beta_1\xi_{t-1,i}+\theta_t$, $\theta_t \stackrel{iid}{\sim} N(0,1)$, and $\{\xi_{t,i},i = 1,\ldots, 1000\}$ are independent of our response. 
We set $\beta_1=.4$, and estimate:
\begin{equation} \hat{\bm{\Upsilon}}=(\hat{\Upsilon}_1,\ldots,\hat{\Upsilon}_{1000}), \textrm{ where } \hat{\Upsilon}_k=\widehat{pdcor}(Y_t,\xi_{t,k};Y_{t-1},Y_{t-2},Y_{t-3}).\end{equation}
We then set the 99th percentile of $\bm{\hat{\Upsilon}}$ as our estimated threshold parameter $\hat{\Gamma}_n$. In order to avoid conditioning on too many variables, following \cite{weng2017regularization}, we set an upper bound of $\lceil n^{1/2} \rceil$ variables which can be added to our conditioning vector at each lag level. Given that $Y_t$ and $\xi_{i,t}$ are independent, we have that $\hat{\Upsilon}_i$ is a statistical estimate of zero. This procedure is similar to the random decoupling approach used in \cite{weng2017regularization} and \cite{Fan2016} for the iid setting. We note that we could also use a more computationally intensive alternative approach which involves forming a pseudo-sample $\{Y_i,\bm{z}_{i}^{*}\}_{i=1}^{n}$, where $(\bm{z}_1^{*},\ldots,\bm{z}_n^{*})$ is formed using a moving block bootstrap \citep{kunsch1989} or a stationary bootstrap \citep{Politis94}. We then use the procedure outlined in \cite{Fan2016}, using the pseudo-sample $\{Y_i,\bm{z}_{i}^{*}\}_{i=1}^{n}$, to select our threshold. We choose to go with the first approach due to computational efficiency.

\section{Screening for Multivariate Time Series Models}\label{section4}

Multivariate time series models, such as linear VAR models, are commonly used in fields such as macroeconomics \citep{Lutkepohl}, finance, and more recently neuroscience \citep{valdes2005}, and genomics. VAR models provide a convenient framework for forecasting, investigating Granger causality, and modeling the temporal and cross-sectional dependence for large numbers of series. Since the number of parameters grows quadratically with the number of component series, VAR models have traditionally been restricted to situations where the number of component series is small. One way to overcome this limitation is by assuming a sparse structure in our VAR process, and using penalized regression methods such as the Lasso and adaptive Lasso \citep{zou2006adaptive} to estimate the model. Examples of works which pursue this direction include \cite{basu2015}, \cite{basu15}, \cite{Kock2015}, and \cite{Matteson16}. However, due to the quadratically increasing nature of the parameter space, penalized regression methods can quickly become computationally burdensome when we have a large panel of component series. For example, in a VAR($k$) process: $\bm{x}_t= \sum_{i=1}^{k}B_i\bm{x}_{t-i} + \bm{\eta}_t$, where $\bm{x}_t \in \mathcal{R}^{m_n}$, $m_n=1000, k=5$, the number of parameters to estimate is $5 \times10^6$. Additionally, these methods are restricted to linear VAR models, whereas there is considerable evidence of non-linear effects such as the existence of thresholds, smooth transitions, regime switching, and varying coefficients in fields such as macroeconomics and finance \citep{Kilian2017}.  

Screening approaches can be used in this setting, and one option would be to screen separately for each of the $m_n$ series. This can be computationally prohibitive since it requires estimating $km_n^2$ correlations. However, if we assume a group structure in the component series and a sparse conditional dependency structure between these groups, we can quickly reduce the feature space by screening at the group level using distance correlation based methods. To be more precise, let $\bm{x}_t$ be a non-linear VAR($k$) process: 
\begin{equation} \bm{x}_t=g(\bm{x}_{t-1},\ldots,\bm{x}_{t-k})+\bm{\eta}_t, \textrm{ where } \bm{x}_t \in \mathcal{R}^{m_n}, \bm{\eta}_t \textrm{ } iid. \end{equation}
For simplicity, we let all groups be of size $g_n$, let $e_n=m_n/g_n$ denote the total number of groups for a given lag level, and denote our groups $(G_{t-1,1},\ldots,G_{t-k,e_n})$. To get a sense of the computational benefits of screening on the group level, assume for example, $m_n=500, k=1$, and we have 25 groups all of size $g_n=20$. 
For this linear VAR $(1)$ model, when $n=200$, we note it takes about 350 times longer to compute all $m_n^2=500^2$ pairwise distance correlations $\left\{\widehat{dcor}(X_{t,j},X_{t-1,k})\right\}_{j\leq m_n,k \leq m_n}$ vs. computing all $e_n^2=25^2$ group pairwise distance correlations. After the group screening, examples of second stage procedures include: screening at the individual series level using partial distance correlations, or using a group lasso type procedure \citep{yuan2006model} which can handle sparsity between groups and within groups for a linear VAR model \citep{basu15}.

We now present the details of our group PDC-SIS procedure. We decide to  condition on only one lag of the grouped response in our procedure, however this number can also be selected using a data driven procedure. Let \\$\mathcal{A}^{(i)}=\left\{(i,k,j): k \in \left\{t-1,
\ldots,t-h\right\},j \leq e_n\right\} \setminus (i,t-1,i)$, refer to the set of possible group connections for $G_{t,i}$. We remove the entry $(i,t-1,i)$ from $\mathcal{A}^{(i)}$, since we are conditioning on $G_{t-1,i}$ and it will not be screened. Let the active group connections for group $i$ be denoted as:
\begin{align*}\mathcal{M}_*^{(i)}=\Bigg\{(i,k,j)\in \mathcal{A}^{(i)}: F\left(G_{t,i}|G_{t-1,i},\bigcup\limits_{r=t-h}^{t-1}\left\{G_{r,l}\right\}_{ l \leq e_n}\right)\textrm{ functionally depends on } G_{k,j} \Bigg\}.\end{align*}
Now let the overall active group connections set be denoted as $\mathcal{M}_*=\bigcup\limits_{i=1}^{m_n}\mathcal{M}_*^{(i)}$. Similarly, our overall screened set is now:
\begin{align*}\hat{\mathcal{M}}_{\gamma_n}=\bigcup\limits_{i=1}^{e_n}\hat{\mathcal{M}}_{\gamma_n}^{(i)}=\left\{(i,k,j) \in \bigcup\limits_{i=1}^{e_n}\mathcal{A}^{(i)} : |\widehat{pdcor}(G_{t,i},G_{k,j};G_{t-1,i})| \geq \gamma_n,\right\}.\end{align*}
The sure screening properties of our group PDC-SIS procedure are similar to the ones presented in theorem 1, and are presented in the supplementary material. From these results, we can infer the maximum size of the groups is $o(n^{1/2-\kappa})$. Given this bound on the group size, our group PDC-SIS procedure is most advantageous when the number of component series ($m_n$) increases polynomially with the sample size. This is usually the case in most VAR models seen in practice. A group version of PDC-SIS+ can also be developed similarly to the procedure in section 3, however we do not pursue this direction, as it usually leads to situations where we are conditioning on large numbers of variables.

\section{Simulations}
\label{section5}
\subsection{Univariate response models: PDC-SIS}
\label{sec5.1}
In the first two subsections, we evaluate the performance of PDC-SIS and PDC-SIS+. We also include the performance of 4 other screening methods whose properties have been investigated in the time series setting, these include: marginal correlation screening (SIS), nonparametric independence screening (NIS), generalized least squares screening (GLSS), and distance correlation screening (DC-SIS). The NIS estimator is computed using the R package \textbf{mgcv}, and the distance and partial distance correlation estimators are computed using the R package \textbf{energy}. For computational efficiency, the GLSS estimator is computed using the \textbf{nlme} package using an AR(1) approximation for the residual covariance matrix. Simulations for our group PDC-SIS procedure are contained in the supplementary material.

Unless noted otherwise, we fix our sample size $n=200$, maximum number of lags considered $h=3$, and the conditioning vector always includes three lags of our response. We vary the number of candidate series, $m_n$, from 500 to 1500, so the number of total covariates, $p_n$, varies from 1500 to 4500. We repeat each experiment 200 times, and report the median minimum model size needed to include all the relevant covariates from $\bm{z}_{t-1}=(\bm{x_{t-1}},\ldots,\bm{x}_{t-3})$. We note that for all procedures being considered, we will not be screening the lags of $Y_t$. In the supplementary materials, we also report the median rank of our relevant covariates for each procedure. We set $Y_0=Y_{-1}=\ldots= Y_{-(h+1)}=0$, and generate $n+200$ samples of our model. We then discard the first $200-h$ samples. To ensure stationarity when generating a nonlinear autoregressive model with exogenous predictors (NARX), we use the sufficient conditions provided in \cite{Masry97}. 

\noindent\textbf{Model 1:}
\begin{align}
Y_t=\sum_{j=1}^{6}\beta_j X_{t-1,j}+\epsilon_t, \textrm{and } \bm{x}_t=A_1\bm{x}_{t-1} + \bm{\eta}_t, \label{VAR1}
\end{align}
where  $A_1=.6* I$, and $\bm{\eta}_t \stackrel{iid}{\sim} N(0,\Sigma_{\eta})$, or $\bm{\eta}_t \stackrel{iid}{\sim} t_{5}(0,3/5*\Sigma_{\eta})$. For this model, we set $\Sigma_{\eta}=\{.3^{|i-j|}\}_{i,j \leq m_n}$. For the error process, we have an AR(1) process: $\epsilon_t=\alpha\epsilon_{t-1}+e_t$ where $\alpha=.6$, and let $e_t \stackrel{iid}{\sim} N(0,1)$ or $e_t \stackrel{iid}{\sim} t_{5}$. This is a linear model with no autoregressive terms, therefore our conditional set contains irrelevant predictors which are the 3 lags of $Y_t$.

The results are displayed in table \ref{table1}, and the entries below ``Gaussian" correspond to the setting where both $e_t$ and $\bm{\eta}_t$ are drawn from a Gaussian distribution. Accordingly the entries under ``$t_5$" correspond to the case where $e_t$ and $\bm{\eta}_t$ are drawn from a $t_5$ distribution. We see that all methods perform well in this scenario, with GLSS performing best and  PDC-SIS following closely even though the lags of $Y_t$ are not significant variables in this example. The results also show that the effects of heavy tails deteriorates the performance of all methods for this model. 

\noindent\textbf{Model 2:}
\begin{align*}
Y_t=g_1(Y_{t-1})+g_2(Y_{t-2})+g_3(Y_{t-3})+f_1(X_{t-1,1})+f_2(X_{t-2,1})+f_3(X_{t-1,2})+f_4(X_{t-2,2})+\epsilon_t, \end{align*}
where the functions are defined as:
\begin{align*}&g_1(x)=.25x,\textrm{ } g_2(x)=x\exp(-x^2/2),\textrm{ } g_3(x)=-.6x+.3x(x>0),\\
 &f_1(x)=1.5x+.4x(x>0),\textrm{ } f_2(x)=-x,\textrm{ } f_3(x)=1.2x+.4x(x>0),\\
 &f_4(x)=x^2sin(2\pi x).\end{align*}
The covariate process is generated as in (\ref{VAR1}), with $A_1=\{.4^{|i-j|+1}\}_{i,j \leq m_n}$ and we set $\Sigma_{\eta}=I_{m_n}$ with $\bm{\eta}_t \stackrel{iid}{\sim} \textrm{ }N(0,\Sigma_{\eta})$ or $\bm{\eta}_t \stackrel{iid}{\sim} t_{5}(0,3/5*\Sigma_{\eta})$. Additionally, we set $\epsilon_t \stackrel{iid}{\sim} \textrm{ }N(0,1)$ or $~t_5$. The nonlinear transformations used are mainly threshold or smooth threshold functions which are popular nonlinear transformations for time series data \citep{Granger2011}. Note that when $x$ is close to zero, $g_2(x)=x$, and when $|x|$ is far from zero the function is close to zero. The results are shown in table \ref{table1}, and our method clearly outperforms the other methods across all scenarios. As seen in table 4 of the supplementary file, the covariate $X_{t-2,1}$ seems to be the most difficult to detect for the competing methods, and it appears our conditioning scheme greatly improves the detection of this signal. 

\noindent\textbf{Model 3:}
\begin{align*}
Y_t&=g_1(Y_{t-1})+g_2(Y_{t-2},Y_{t-1})+g_3(Y_{t-3},Y_{t-1})+f_1(X_{t-1,1},X_{t-1,4})\\
&+f_2(X_{t-2,1},X_{t-1,4})+f_3(X_{t-1,2},X_{t-1,4})+f_4(X_{t-2,2},X_{t-1,4})\\
&+f_5(X_{t-1,3},X_{t-1,4})+f_6(X_{t-1,4})+f_{7}(X_{t-1,3},X_{t-1,4})+\epsilon_t, \end{align*}
where the functions are defined as:
\begin{align*}&g_1(x)=.2x+.2x(x>0),\textrm{ } g_2(x,y)=.2x+.1x(y>0),\\
&\textrm{ } g_3(x,y)=x\exp(-y^2/2), f_1(x,y)=f_3(x,y)=f_5(x,y)=x(1+\exp(-y^2/2)),\\
&f_2(x,y)=f_4(x,y)=(1+.5\exp(-y^2/2))x,\textrm{ } f_6(x)=x\exp(-x^2/2), f_7(x,y)=xy.
\end{align*}
The covariate process is a VAR(2) process: $\bm{x}_t=A_1\bm{x}_{t-1} + A_2\bm{x}_{t-2}+\bm{\eta}_t$, where  $A_1=\{.3^{|i-j|+1}\}_{i,j \leq m_n}$, $A_2=\{.2^{|i-j|+1}\}_{i,j \leq m_n}$, and $\Sigma_{\eta}=\{-.3^{|i-j|}\}_{i,j \leq m_n}$. As before, $\bm{\eta}_t \stackrel{iid}{\sim} \textrm{ }N(0,\Sigma_{\eta})$ or $\bm{\eta}_t \stackrel{iid}{\sim} t_{5}(0,3/5*\Sigma_{\eta})$.

In this model the threshold variable for the autoregressive terms is $Y_{t-1}$, while the threshold variable for the covariates is $X_{t-1,4}$. For the covariates we apply a smooth threshold function, and for the autoregressive terms we mainly employ a hard threshold function. Threshold transformations with a single threshold variable imply a change in that variable causes a shift in the effects of all  other covariates, and many examples of this effect can be found in macroeconomics. 

The results are displayed in table \ref{table1}, and our method outperforms the rest of the methods, with distance correlation usually taking second place. The median ranks of each of our significant variables can be found in table 5 of the supplementary file. And the variable which appears to be the most difficult to detect seems to be the threshold variable, $X_{t-1,4}$. As predicted by the theoretical results and inline with results for the previous models, the performance of all  methods worsens as we encounter heavy tails.

\noindent\textbf{Model 4:}
\begin{align*}
Y_t&=.25Y_{t-1}+.3Y_{t-2}+.3Y_{t-3}+f_1(X_{t-1,1})+f_2(X_{t-2,1})\\
&+\beta_{1,t}f_{3}(X_{t-1,2},X_{t-1,3})+\beta_{2,t}f_{4}(X_{t-2,2},X_{t-2,3})+\beta_{3,t}f_5(X_{t-1,3})\\
&+\beta_{4,t}f_6(X_{t-2,3})+f_7(X_{t-1,2})+f_8(X_{t-2,2},X_{t-1,2})+\epsilon_t
,\end{align*}
where the functions are defined as:
\begin{align*}
&f_1(x),f_7(x)=1.5x+.4x(x>0),f_2(x)=1.2x,\textrm{ }f_3(x,y)=f_4(x,y)=xy,\\
&f_5(x),f_6(x)=x \textrm{, }f_8(x,y)=1.2x+.4x(y>0)\textrm{, }\beta_{1,t},\beta_{2,t},\beta_{3,t},\beta_{4,t} \stackrel{iid}{\sim}Unif(.5,1) \textrm{ }\forall t.
\end{align*}
The covariate process is generated as in (\ref{VAR1}), with $A_1=\{.4^{|i-j|+1}\}_{i,j \leq m_n}$ and $\Sigma_{\eta}=\{-.3^{|i-j|}\}_{i,j \leq m_n}$. As in the previous examples, $\bm{\eta}_t \stackrel{iid}{\sim} \textrm{ }N(0,\Sigma_{\eta})$ or $\bm{\eta}_t \stackrel{iid}{\sim} t_{5}(0,3/5*\Sigma_{\eta})$. We also note that the coefficients $\beta_{1,t},\beta_{2,t},\beta_{3,t},\beta_{4,t}$, are random at each time $t$. 
In this model the autoregressive portion is linear, and for the exogenous covariates we use a mix of threshold functions, interactions, and random coefficients. The results are displayed in table \ref{table1}, and our method again does better than the rest of the methods considered, as the others seem to have difficulty dealing with the combination of high dependence and nonlinearities. Looking at table 6 in the supplementary file, we notice that the covariates $X_{t-1,3},X_{t-2,3}$, which only appear through random coefficient effects, are the most difficult to predict. 

\subsection{Univariate response models: PDC-SIS+}
\label{sec5.2}

The previous subsection showed a range of nonlinear time series models in which the performance of PDC-SIS is superior to competing methods. In this subsection, we will compare the performance of PDC-SIS vs. PDC-SIS+, and investigate whether we can improve upon the performance of PDC-SIS by adding variables to our conditioning vector in a data driven way.

Looking at the performance of our previous simulations, we observe that in model 2 we have difficulty detecting $X_{t-2,1}$. Therefore, we start by comparing PDC-SIS vs. PDC-SIS+ on model 2. The results are displayed in table \ref{table5}, and we report the median rank of $X_{t-1,2}$ as well as the median minimum model size (MMS) for both procedures. The results show that PDC-SIS+ clearly outperforms PDC-SIS in detecting $X_{t-2,1}$, and therefore has a much smaller MMS. This difference is seen clearest in the case where we have heavy tails and $p_n=4500$. The median MMS for PDC-SIS+ is 125 vs. 275 for PDC-SIS.

Our next model is a linear ARDL model:\\
\noindent\textbf{Model 5:} $Y_t=.25Y_{t-1}+.3Y_{t-2}+.3Y_{t-3}+X_{t-1,1}-X_{t-2,1}+.5X_{t-1,2}+.5X_{t-2,2}+\epsilon_t.$
\\

The covariate process is generated as in (\ref{VAR1}), with $A_1=\{.4^{|i-j|+1}\}_{i,j \leq m_n}$ and we set $\Sigma_{\eta}=I_{m_n}$ with $\bm{\eta}_t \stackrel{iid}{\sim} \textrm{ }N(0,\Sigma_{\eta})$ or $\bm{\eta}_t \stackrel{iid}{\sim} t_{5}(0,3/5*\Sigma_{\eta})$. Additionally we set $\epsilon_t \stackrel{iid}{\sim} \textrm{ }N(0,1)$ or $t_5$. By construction $X_{t-2,1}$ would be the most difficult to detect using a marginal approach. The results are in table \ref{table6}, and show that PDC-SIS+ does better in all scenarios, with the difference being most pronounced in the case of heavy tails. The results from both model 2 and model 6 suggest that adding strong conditional signals to our conditioning vector improves upon the performance of PDC-SIS.

\begin{table}[]
\footnotesize
\centering
\caption{Median Minimum Model Size}
\label{table1}
\begin{tabular}{|c|c|c|c|c|}
\hline
\multicolumn{5}{|c|}{Gaussian, $p_n=1500$}       \\ \hline
        & Model 1 & Model 2 & Model 3 & Model 4 \\ \hline
PDC-SIS & 7       & 61      & 29      & 42      \\ \hline
DC-SIS  & 11      & 488     & 112     & 306.5   \\ \hline
NIS     & 11      & 488     & 119.5   & 275     \\ \hline
SIS     & 10      & 343.5   & 100.5   & 234.5   \\ \hline
GLSS    & 6       & 179.5   & 813     & 800.5   \\ \hline
\end{tabular}
\begin{tabular}{|c|c|c|c|c|}
\hline
\multicolumn{5}{|c|}{Gaussian, $p_n=4500$}       \\ \hline
        & Model 1 & Model 2 & Model 3 & Model 4 \\ \hline
PDC-SIS & 11      & 149     & 78.5    & 100.5   \\ \hline
DC-SIS  & 19      & 1051    & 337     & 842.5   \\ \hline
NIS     & 16      & 861     & 309     & 704     \\ \hline
SIS     & 13      & 722     & 281     & 588     \\ \hline
GLSS    & 6       & 592     & 2325.5  & 2214    \\ \hline
\end{tabular}
\begin{tabular}{|c|c|c|c|c|}
\hline
\multicolumn{5}{|c|}{$t_5$, $p_n=1500$}          \\ \hline
        & Model 1 & Model 2 & Model 3 & Model 4 \\ \hline
PDC-SIS & 13      & 79.5    & 43      & 51      \\ \hline
DC-SIS  & 20      & 408.5   & 114     & 306     \\ \hline
NIS     & 33      & 513.5   & 167     & 328     \\ \hline
SIS     & 21.5    & 447     & 166.5   & 265     \\ \hline
GLSS    & 6       & 450.5   & 969.5   & 891.5   \\ \hline
\end{tabular}
\begin{tabular}{|c|c|c|c|c|}
\hline
\multicolumn{5}{|c|}{$t_5$, $p_n=4500$}          \\ \hline
        & Model 1 & Model 2 & Model 3 & Model 4 \\ \hline
PDC-SIS & 36.5    & 275.5   & 78      & 104     \\ \hline
DC-SIS  & 68      & 951.5   & 301.5   & 814.5   \\ \hline
NIS     & 114     & 1100.5  & 436.5   & 851.5   \\ \hline
SIS     & 66.5    & 905     & 438     & 761     \\ \hline
GLSS    & 7       & 1386.5  & 3008    & 2843.5  \\ \hline
\end{tabular}
\end{table}
\begin{table}[]
\footnotesize
\centering
\caption{Model 2, PDC-SIS+}
\label{table5}
\begin{tabular}{|c|c|c|c|c|}
\hline
                     & \begin{tabular}[c]{@{}c@{}}PDC-SIS+\\ MMS\end{tabular} & \begin{tabular}[c]{@{}c@{}}PDC-SIS+\\ $X_{t-2,1}$\end{tabular} & \begin{tabular}[c]{@{}c@{}}PDC-SIS\\ MMS\end{tabular} & \multicolumn{1}{l|}{\begin{tabular}[c]{@{}l@{}}PDC-SIS\\ $X_{t-2,1}$\end{tabular}} \\ \hline
Gaussian, $p_n=1500$ & 34                                                     & 26                                                           & 61                                                    & 40.5                                                                                 \\ \hline
Gaussian, $p_n=4500$ & 79                                                   & 43.5                                                           & 149                                                   & 141                                                                                \\ \hline
$t_5$, $p_n=1500$    & 57.5                                                     & 29                                                             & 79.5                                                  & 57.5                                                                               \\ \hline
$t_5$, $p_n=4500$    & 121.5                                                  & 88                                                           & 275.5                                                 & 239.5                                                                              \\ \hline
\end{tabular}
\end{table}
\begin{table}[]
\centering
\caption{Model 5, PDC-SIS+}
\footnotesize
\label{table6}
\begin{tabular}{|c|c|c|c|c|}
\hline
                     & \begin{tabular}[c]{@{}c@{}}PDC-SIS+\\ MMS\end{tabular} & \begin{tabular}[c]{@{}c@{}}PDC-SIS+\\ $X_{t-2,1}$\end{tabular} & \begin{tabular}[c]{@{}c@{}}PDC-SIS\\ MMS\end{tabular} & \multicolumn{1}{l|}{\begin{tabular}[c]{@{}l@{}}PDC-SIS\\ $X_{t-2,1}$\end{tabular}} \\ \hline
Gaussian, $p_n=1500$ & 22                                                  & 22                                                             & 24                                                    & 23                                                                                 \\ \hline
Gaussian, $p_n=4500$ & 42                                                     & 40.5                                                           & 59                                                    & 58                                                                                 \\ \hline
$t_5$, $p_n=1500$    & 42                                                   & 39                                                             & 52                                                    & 52                                                                                 \\ \hline
$t_5$, $p_n=4500$    & 85                                                     & 76.5                                                           & 162.5                                                 & 159.5                                                                              \\ \hline
\end{tabular}
\end{table}

\section{Real Data Application: Macroeconomic Forecasting}
\label{section6}

In this section, we present an application to forecasting univariate macroeconomic time series. An application to multivariate time series models can be found in the supplementary material. Our dataset consists of 132 monthly macroeconomic series which run from January 1984 to December 2011, and  was obtained from the supplement to \cite{Jurado15}. The transformations needed to achieve approximate stationarity as well as descriptions of the series are given in \cite{Jurado15}. A start date of January 1984 is chosen since most macroeconomic series are widely thought to contain a structural break around the first quarter of 1984, resulting in significantly lower volatility in the U.S economy \citep{SW2003}. This effect has been known as the great moderation in macroeconomics, and various explanations for this phenomenon are given in \cite{Boivin06,SW2003}. 

We focus on forecasting the 6 month ahead real personal income less transfer payments (RPI), and the 6 month ahead number of employees
on nonfarm payrolls (EMP). Both of these series are major monthly economic series which are closely watched by the National Bureau of Economic Research (NBER) business cycle dating committee \citep{NBER}. 
We utilize a rolling window scheme, where the first simulated out of sample forecast was for the time period 2000:1 (January 2000). To construct this forecast, we use the observations between 1984:4 to 1999:7 (the first three observations are used in forming lagged covariates) to estimate the factors, and the coefficients. Therefore for the models described above, $t=$1984:4 to 1999:1. We then use the regressor values at $t=$1999:7 to form our forecast for 2000:1. The next window uses observations from 1984:5 to 1999:8 to forecast 2000:2. Using this scheme we have a total of 144 out of sample forecasts, and for each window we use $n=178$ observations.

We assume the following model for our forecasts:
\begin{equation} Y_{t+6}^6= g(Y_t,\ldots,Y_{t-3},\bm{z}_t)+\epsilon_{t+6}^6\label{predmodel}. \end{equation} 
where $Y_{t+6}^{6}=\log(RPI_{t+6}/RPI_{t})$, and $Y_t=\log(RPI_t/RPI_{t-1})$, and we replace RPI with EMP in the previous definitions when forecasting EMP. Additionally, $\bm{z}_t=(\bm{x}_t,\ldots,\bm{x}_{t-3})$, where $\bm{x}_t$ are the 131 macroeconomic series apart from $y_t$, which gives us a total of 528 predictors. We report the forecasting performance of 15 different models. The first is a baseline linear AR(4) model: $\hat{Y}_{t+6}^{6}=\hat{\alpha}_0+\sum_{i=0}^{3} \hat{\alpha}_i Y_{t-i}$. We then combine each of the six screening methods under consideration with two classes of second stage procedures. The first class of models assumes $g(\cdot)$ is a linear model, which we will estimate with the adaptive Lasso. For the adaptive Lasso, we use the Lasso as our initial estimate, and we choose the penalty parameters for both the adaptive Lasso and the Lasso using the modified BIC \citep{Wangetal2009}. The second class of models are \emph{factor augmented autoregressions}: $\hat{Y}_{t+6}^{6}=\hat{\beta_0}+\sum_{i=0}^{3} \hat{\alpha}_i Y_{t-i}+\hat{\bm{\gamma}} \bm{\hat{F}}_{t}$, where $\bm{\hat{F}}_{t}=(\hat{F}_{t,1},\ldots,\hat{F}_{t,4})$ are four factors which are computed as the first four principal components of the top $d_n^{'}= n-1$ predictors of $\bm{z}_t$, as ranked by the screening procedures. Lastly, we include the performance of both classes of models on the entire dataset of 528 predictors.

For PDC-SIS and PDC-SIS+, $(Y_t,\ldots,Y_{t-3})$ is part of the conditional vector for each variable, and we compute each of the screening methods as discussed in section \ref{section5}. When using the adaptive Lasso as a second stage method, we select the top $d_n=\lceil n/\log(n) \rceil=35$ predictors from $z_t$ as our screened set for all screening methods. We then add the vector $(Y_t,\ldots,Y_{t-3})$ to our screened set and estimate our model via the adaptive Lasso using this subset of predictors.

The results are reported in table \ref{table9}, and the entries in bold refer to the best performing model(s). We report the mean squared error (MSE), and the mean absolute error (MAE) of the resulting forecasts relative to the MSE, MAE of the benchmark AR(4) forecasts respectively. For RPI, when using adaptive Lasso as our second stage method we generally see distance correlation and PDC based screening algorithms outperforming competing procedures across both error measures. For employment forecasting, using the adaptive Lasso as a stand alone or second stage procedure appears to do worse than the baseline AR(4) model. For both RPI and Employment,  factor augmented autoregressions clearly outperform the baseline AR(4) and adaptive lasso forecasts for both error measures. For both series our PDC based factor forecasts outperform competing screening methods with the difference being larger when forecasting employment. It is also encouraging to see that our PDC based factor forecasts outperform the standalone factor forecasts. This is inline with results reported in \cite{Ng2006}, which showed that adding too many irrelevant variables can deteriorate its forecasting performance. We note that using a model free screening procedure gives us full flexibility in choosing a second stage procedure, so our results might be further improved by considering additional classes of second stage procedures.

\begin{table}[]
\small
\centering
\caption{\small{Employment (EMP) and Real Personal Income (RPI): 6 month ahead forecasts}}
\label{table9}
\begin{tabular}{|c|c|c|c|c|}
\hline
                       & \multicolumn{2}{c|}{EMP}        & \multicolumn{2}{c|}{RPI}        \\ \hline
                       & MSE            & MAE            & MSE            & MAE            \\ \hline
AR (4)                 & 1.00           & 1.00           & 1.00           & 1.00           \\ \hline
SIS-Adaptive Lasso     & 1.22           & 1.05           & .77            & .92            \\ \hline
GLSS-Adaptive Lasso    & .89            & .98            & .78            & .91            \\ \hline
NIS-Adaptive Lasso     & 1.23           & 1.08           & .83            & .95            \\ \hline
DC-SIS Adaptive Lasso  & 1.15           & 1.04           & .71            & .88            \\ \hline
PDC-SIS Adaptive Lasso & 1.19           & 1.06           & .72            & .89            \\ \hline
PDC-SIS+Adaptive Lasso & 1.24           & 1.07           & .73            & .91            \\ \hline
Adaptive Lasso         & 1.14           & 1.05           & .80            & .96            \\ \hline
SIS-Factor AR          & .83            & .84            & .56            & .77            \\ \hline
GLSS-Factor AR         & .82            & .87            & .66            & .83            \\ \hline
NIS-Factor AR          & .91            & .86            & .62            & .81            \\ \hline
DC-SIS Factor AR       & .78            & .81            & .57            & .78            \\ \hline
PDC-SIS Factor AR      & \textbf{.63} & \textbf{.77} & \textbf{.52}            & \textbf{.75}            \\ \hline
PDC-SIS+ Factor AR     & .68            & .79            & \textbf{.52} & \textbf{.75} \\ \hline
Factor AR              & .70            & .81            & .58            & .81            \\ \hline
\end{tabular}
\vspace{1ex}

\footnotesize{\textit{MSE, MAE reported are relative to the benchmark AR(4) forecasts}}
\end{table}

\section{Discussion}
\label{section7}

In this work, we have introduced two classes of partial distance correlation based screening procedures, which are applicable to univariate or multivariate time series models. These methods aim to utilize the unique features of time series data as an additional source of information, rather than treating temporal dependence as a nuisance. By using a model free first stage procedure we are able to expand the choice of models which can be considered for a second stage procedure. This is especially helpful for the case of nonlinear or nonparametric models where estimation in high dimensions can be computationally challenging. 

There are many opportunities for further research, such as developing a theoretical or data driven approach to selecting the number of lags considered in our algorithms. Additionally, we can develop screening algorithms for time series data using measures which are more robust to heavy tailed distributions. Lastly, our procedures were developed under the assumption that the underlying processes are weakly dependent and stationary. Although these assumptions are satisfied for a very wide range of applications, there are many instances where they are violated. For example, non-stationarity is commonly induced by time varying parameters, structural breaks, and cointegrated processes, all of which are common in the fields of macroeconomics and finance. In addition, long range dependence is a property which is prominent in economics, finance, climate studies, and the physical sciences (see \cite{Samorodnitsky2006} for more details). Therefore, developing new methodologies for long range dependent processes, or certain classes of non-stationary processes, such as locally stationary processes, would be particularly welcome. 

\section*{Supplementary Material}
Due to space limitations, simulations and a real data application of our group PDC-SIS procedure are contained in the supplementary material. Additionally, the supplementary material also contains the proofs for all the theorems, as well as more detailed tables of the results from section \ref{sec5.1}.

\begin{spacing}{0.5}
\bibliographystyle{biom}
{\small
\bibliography{varselection}}
\end{spacing}

\newpage
\setcounter{page}{1}
\setcounter{section}{1}
\setcounter{table}{0}
\renewcommand{\theequation}{A.\arabic{equation}}
\setcounter{equation}{0}

\section*{\bf Supplement to Partial Distance Correlation Screening for High Dimensional Time Series}

\maketitle

This supplementary document is organized as follows: supplement A contains the sure screening properties, simulations, as well as a real data application of our group PDC-SIS procedure. Supplement B contains the proofs of theorems 1 and 2 found in our main paper. Lastly supplement C provide more detailed results of the simulations in section 5.1 of our main paper.

\section*{Supplement A}
\subsection{Sure Screening Properties for Group PDC-SIS}
As in our main paper, we assume the multivariate response process has the representation:
\begin{align}\bm{x_i}=\bm{h}\left(\ldots,\bm{\eta}_{i-1},\bm{\eta}_i\right). \label{nonlinear2}\end{align} 
Where $\bm{\eta}_i, i\in \mathbb{Z}$, are iid random vectors. To prove sure screening properties of our group PDC-SIS procedure, we need the following conditions:
\begin{condition}\label{ConditionF} 
Assume $|pdcor(G_{t,i},G_{k,j};G_{t-1,i})| \geq c_1 n^{-\kappa}$ for $(i,k,j) \in M_* \textrm{ , }\kappa \in (0,1/2).$
\end{condition}
\begin{condition}\label{ConditionB}
Assume our multivariate response process has the representation (\ref{nonlinear2}). Additionally, we assume the following decay rate $\Phi_{m,r}(\bm{x}) = O(m^{-\alpha_x})$, for some $\alpha_x > 0$, $r > 4 $. 
\end{condition}
\begin{condition}\label{ConditionC} 
Assume our multivariate response process $\bm{x_t}$ has the representation (\ref{nonlinear2}). Additionally assume $\upsilon_z=\sup_{q \geq 2} q^{-\tilde{\alpha}_x}\Phi_{0,q}(\bm{x}) < \infty$, for some $\tilde{\alpha}_x \geq 0.$
\end{condition}
\begin{condition}\label{ConditionD}
Assume the process $\left\{\bm{x}_t\right\}$ is $\beta$-mixing, with mixing rate $\beta_{x}(a)=O(\exp(-a^{\lambda_1}))$, for some $\lambda_{1} >0$.
\end{condition}

Let $\varrho =1$, if $\alpha_x > 1/2 - 2/r$, otherwise $\varrho= r/4 - \alpha_x r/2$. And let $K_{x,r}=\sup_{m \geq 0} (m+1)^{\alpha_x} \Phi_{r}(\bm{x})$. Recall that $t_n=\max_j dim(C_j)$ is the maximum dimension of the conditional vectors. Lastly let $\tilde{\varphi} = \frac{2}{1+4\tilde{\alpha}_x}$. The results are similar to those in theorem 1, but for the sake of completeness we present them here as well:
\begin{cor}   
\begin{enumerate}[noitemsep]
\item Suppose conditions  \ref{ConditionF}, \ref{ConditionC}, \ref{ConditionD} hold. For $\gamma_n = c_3 n^{-\kappa}$ with $c_3 \leq c_1/2$, we have:
\begin{eqnarray*} P\left(\mathcal{M}_*\subset \hat{\mathcal{M}}_{\gamma_n} \right)\geq 1-O\left(s_nn^2\exp\left(-\frac{n^{1/2-\kappa}}{t_n\upsilon_z^2}\right)^{\tilde{\varphi}}\right).
\end{eqnarray*}
\item Suppose conditions \ref{ConditionF}, \ref{ConditionB}, \ref{ConditionD} hold. For $\gamma_n = c_3 n^{-\kappa}$ with $c_3 \leq c_1/2$, we have
$$\mbox{ if $r<12$,}\quad P\left(\mathcal{M}_*\subset \hat{\mathcal{M}}_{\gamma_n} \right)\geq 1-O(s_n\frac{t_n^{r/2}K_{x,r}^{r}}{n^{r/4-r\kappa/2}});$$
$$\mbox{ if $r\ge 12$,}\quad P\left(\mathcal{M}_*\subset \hat{\mathcal{M}}_{\gamma_n} \right)\geq 1-O\left(s_nn^2\Bigg[\frac{t_n^{r/2}n^{\varrho} K_{x,r}^{r}}{n^{r/2-r\kappa/2}}+\exp\left(-\frac{n^{1-2\kappa}}{t_n^2K_{x,r}^4}\right)\Bigg]\right).$$
\end{enumerate}
\end{cor}
From the above results we can infer the maximum size of the groups is $o(n^{1/2-\kappa})$. The proof for this corollary is very similar to the proof of theorem 1, therefore we omit the details.

\subsection{Simulations: Group PDC-SIS}
\label{sec5.3}
We consider the following VAR(1) process, \\ \textbf{Model 6}:
\begin{equation}\bm{x}_t=A_1\bm{x}_{t-1} + \bm{\eta}_t, \end{equation}
and  assume we have 25 groups at each lag level ($e_n=25$) with equal size $g_n=20$. We assume a block upper triangular structure for $A_1$, with two scenarios. 
\begin{equation} \label{myeq}
A_1 = \left[ \begin{matrix}
{B} & {0} &{C}&& 0\\
& \ddots & \ddots & \ddots &\\
&  & \ddots &\ddots &{C} \\
& & & \ddots & {0} \\
0 & & & & {B} \end{matrix} \right]
.\end{equation}
We set the number of lags considered, $h=2$, therefore we have to compute 1225 group distance and partial distance correlations for each scenario. In the first scenario we set the main diagonal blocks to $B=\{.3^{|i-j|+1}\}_{i,j \leq g_n}$,  the second upper diagonal blocks to $C=\{.2^{|i-j|+1}\}_{i,j \leq g_n}$, and the rest of the matrix to zero. In the second scenario, we assume the same number of groups and group size, but we set the diagonal group $B=\{.3^{|i-j|+1}\}_{i,j \leq 10}$, and the second upper diagonal block to $C=\{.2^{|i-j|+1}\}_{i,j \leq 10}$. We can view this scenario as one in which we have misspecified the groups \citep{basu15}, or one in which we have sparsity within each group. We set $\Sigma_{\eta}=\{.4^{|i-j|+1}\}_{i,j \leq m_n}$ or $\Sigma_{\eta}=\{-.4^{|i-j|+1}\}_{i,j \leq m_n}$. And lastly, $\bm{\eta}_t \stackrel{iid}{\sim} \textrm{ }N(0,\Sigma_{\eta})$ or $\bm{\eta}_t \stackrel{iid}{\sim} t_{3}(0,1/3*\Sigma_{\eta})$. 

Since we are assuming the first lag for each group is in the model, we have 23 off-diagonal group connections we want to detect for each scenario. As in our main paper, the sample size is $n=200$, and we report the median MMS for group DC-SIS, and group PDC-SIS procedure for each scenario in table \ref{table7}. The MMS in this case is defined as the minimum number of group connections which need to be selected for $\mathcal{M}_{*}$ to be captured. In order to ensure a fair comparison, we do not evaluate $dcor(G_{t,i},G_{t-1,i})$ for each group $i$ when using group DC-SIS. The results show that the procedures are robust to the level of sparsity within each group, and our group PDC-SIS procedure significantly outperforms the group DC-SIS for all scenarios.

\begin{table}[]
\centering
\caption{Model 6}
\label{table7}
\begin{tabular}{|c|c|c|c|c|}
\hline
                                          & \begin{tabular}[c]{@{}c@{}}Scenario 1\\ PDC-SIS\end{tabular} & \begin{tabular}[c]{@{}c@{}}Scenario 1\\ DC-SIS\end{tabular} & \begin{tabular}[c]{@{}c@{}}Scenario 2\\ PDC-SIS\end{tabular} & \begin{tabular}[c]{@{}c@{}}Scenario 2\\ DC-SIS\end{tabular} \\ \hline
$N(0, \Sigma_{\eta}=\{.4^{|i-j|+1}\})$    & 33                                                            & 53                                                           & 32                                                            & 52                                                           \\ \hline
$N(0, \Sigma_{\eta}=\{-.4^{|i-j|+1}\})$   & 68                                                            & 139.5                                                        & 66                                                            & 140                                                          \\ \hline
$t_3$, $\Sigma_{\eta}=\{.4^{|i-j|+1}\})$  & 38                                                            & 46.5                                                         & 37                                                            & 45                                                           \\ \hline
$t_3$, $\Sigma_{\eta}=\{-.4^{|i-j|+1}\})$ & 89                                                            & 159.5                                                        & 83.5                                                          & 145.5                                                        \\ \hline
\end{tabular}
\end{table}

\subsection{Real data application: Group PDC-SIS}

For the multivariate response setting, we focus on the group selection performance. We partition the 132 economic series into 8 broad economic groups: 1) Output and income (17 series) 2) Labor Market (32 series) 3) Housing (10 series) 4) Consumption, Orders, and Inventories (14 series) 5) Money and Credit (11 series) 6) Bonds and Exchange rates (22 series) 7) Prices (21 series) 8) Stock market (4 series). We then supplement this with 300 additional exogenous series ($\bm{v}_t$) partitioned into groups of size 10. Where  $\bm{v}_t=A_1\bm{v}_{t-1} + \bm{\eta}_t$, $A_1=\alpha * I$, where we vary $\alpha$ from .4 to .8, and we $\bm{\eta}_t \stackrel{iid}{\sim}  N(0,I)$ or $\bm{\eta}_t \stackrel{iid}{\sim} t_3(1/3*I)$. We have 38 groups for each lag level, and we set the number of lags considered, $h=2$, giving us about 2900 group comparisons to compute. Let $\bm{x}_t$ represent our 132 economic series, and let $\bm{z}_t=(\bm{x}_t,\bm{v}_t)$ with $\bm{v}_t$ being independent of $\bm{x}_t$. We assume the following one step ahead forecasting strategy:
\begin{equation} \bm{z}_t=\bm{f}(\bm{z}_{t-1},\bm{z}_{t-2})+\bm{\epsilon}_t. \end{equation}
We utilize a rolling window scheme similar to the one described previously, except we are not computing out of sample forecasts. For the first window we use data from $t=$1984:3 to $t=$1999:12 to compute our correlations. We then move the window forward by one month, which gives us 144 windows in total and 191 observations for each window. As discussed in section 4 of our main paper, for each group $G_{t,i}$ we condition on the first lag $G_{t-1,i}$ for PDC-SIS. Let $\left\{G_{t,j}\right\}_{j \leq 8}$ represents the 8 economic groups at time $t$, and let $\mathcal{B}=\left\{(i,k,j):i,j \leq 8, k \in \left\{t-1,t-2\right\}\right\} \setminus \left\{(i,t-1,i): i\leq 8\right\}$ denotes the set of possible group connections between the 8 economic groups minus the connection between a group and its first lag.  For each window, we select the top $\lceil n/\log(n) \rceil=37$ group connections, and record the number of group connections which belong to $\mathcal{B}$. We note that all group connections which are to be screened and do not belong to $\mathcal{B}$ are spurious connections by construction. 

The results are in table \ref{table10}, and we report the median number of group connections which belong to $\mathcal{B}$ over the 144 windows. In order to ensure a fair comparison between group DC-SIS and group PDC-SIS, we do not evaluate $dcor(G_{t,i},G_{t-1,i})$ for each group $i$ when using group DC-SIS. We see that when $\alpha=.4$ and the noise is Gaussian, both group PDC-SIS and group DC-SIS are very effective at selecting connections between economic groups. When the dependence increases and heavy tailed variables are introduced, the performance of group DC-SIS greatly deteriorates with many spurious group connections selected, whereas group PDC-SIS remains effective.

\begin{table}[]
\centering
\caption{Group Selection}
\label{table10}
\begin{tabular}{|c|c|c|}
\hline
                      & PDC-SIS & DC-SIS \\ \hline
Gaussian, $\alpha=.4$ & 37      & 34     \\ \hline
Gaussian, $\alpha=.6$ & 32      & 25     \\ \hline
Gaussian, $\alpha=.8$ & 22      & 9      \\ \hline
$t_3$, $\alpha=.4$    & 36      & 31     \\ \hline
$t_3$, $\alpha=.6$    & 31      & 21.5   \\ \hline
$t_3$, $\alpha=.8$    & 23      & 8      \\ \hline
\end{tabular}
\end{table}

\section*{Supplement B: Proofs of Theorems 1 and 2} \label{sec:SuppA}

\begin{proof}[Proof of Theorem 1]
  $ $\newline
We start with part (iii) first. The population version of the partial distance correlation is defined as:
\begin{equation}pdcor(Y_t,Z_{t-1,k};C_k)=\frac{dcor^2(Y_t,Z_{t-1,k})-dcor^2(Y_t,C_k)dcor^2(Z_{t-1,k},C_k)}{\sqrt{1-dcor^4(Y_{t},C_k)}\sqrt{1-dcor^4(Z_{t-1,k},C_k)}}. \label{pddef}\end{equation}
To estimate this quantity, \cite{Szekeley2014} proposed an unbiased estimator of the distance correlation to serve as the plug-in estimate. This estimate is different from the estimator proposed for the distance correlation in \cite{Szekeley2007}, which is a biased but consistent estimate. In proving asymptotic properties we can use either estimate, and we will use the original estimator given in \cite{Szekeley2007}. 

To obtain a bound for $|\widehat{pdcor}(Y_t,Z_{t-1,k};C_k) -pdcor(Y_t,Z_{t-1,k};C_k)|$, we start with $|\widehat{dcor}^2(Y_t,Z_{t-1,k})-dcor^2(Y_t,Z_{t-1,k})|$ in the numerator of (\ref{pddef}). Recall that:
\begin{equation} \widehat{dcor}^2(Y_{t},Z_{t-1,k})=\frac{\widehat{dcov}^2(Y_{t},Z_{t-1,k})}{\widehat{dcov}(Y_t,Y_t)\widehat{dcov}(Z_{t-1,k},Z_{t-1,k})} \label{dcdef}. \end{equation}
Let $\hat{T}_1=\widehat{dcov}^2(Y_{t},Z_{t-1,k})$,$\hat{T_2}=\widehat{dcov}(Y_t,Y_t)\widehat{dcov}(Z_{t-1,k},Z_{t-1,k})$, and $T_1=dcov^2(Y_{t},Z_{t-1,k})$, $T_2=dcov(Y_t,Y_t)dcov(Z_{t-1,k},Z_{t-1,k})$, then 
\begin{align} |\widehat{dcor}^2(Y_t,Z_{t-1,k})-dcor^2(Y_t,Z_{t-1,k})|=|\frac{\hat{T}_1}{\hat{T}_2}-\frac{T_1}{T_2}| \nonumber \\
=|(\hat{T}_2^{-1}-T_2^{-1})(\hat{T}_1-T_1)+(\hat{T}_1-T_1)/T_2+(\hat{T}_2^{-1}-T_2^{-1})T_1|. \end{align}
Therefore
 \begin{align} P(|\frac{\hat{T}_1}{\hat{T}_2}-\frac{T_1}{T_2}|> cn^{-\kappa})&\leq P(|(\hat{T}_2^{-1}-T_2^{-1})(\hat{T}_1-T_1)|>c_2n^{-\kappa}/3)\label{triple1}\\
 &+P(|(\hat{T}_1-T_1)/T_2|>c_2n^{-\kappa}/3|)\label{triple2}\\
 &+P(|(\hat{T}_2^{-1}-T_2^{-1})T_1|>c_2n^{-\kappa}/3).\label{triple3}
 \end{align}
For the RHS of (\ref{triple1}), we obtain:
\begin{align*} P(|(\hat{T}_2^{-1}-T_2^{-1})(\hat{T}_1-T_1)|>c_2n^{-\kappa}/3) &\leq P(|\hat{T}_1-E(T_1)|>Cn^{-\kappa/2})\\
&+ P(|\hat{T}_2^{-1}-E(T_2)^{-1}|>Cn^{-\kappa/2}).\end{align*}
So we focus on terms (\ref{triple2}) and (\ref{triple3}). For (\ref{triple2}), recall that:
\begin{equation} \hat{dcov^2}(Y_t,Z_{t-1,k})=\hat{S}_{k1} + \hat{S}_{k2} - 2\hat{S}_{k3}, \end{equation}
where
\begin{align} \hat{S}_{k1}=n^{-2}\sum_{j=1}^{n}\sum_{i=1}^{n}|Y_i-Y_j||Z_{i,k}-Z_{j,k}|,\nonumber \\
\hat{S}_{k2}=n^{-2}\sum_{j=1}^{n}\sum_{i=1}^{n}|Y_i-Y_j|n^{-2}\sum_{j=1}^{n}\sum_{i=1}^{n}|Z_{i,k}-Z_{j,k}|, \nonumber \\
\hat{S}_{k3}=n^{-3}\sum_{j=1}^{n}\sum_{i=1}^{n}\sum_{l=1}^{n}|Y_i-Y_j| |Z_{i,k}-Z_{l,k}|. \end{align}
We begin with the term $|\hat{S}_{k1}-S_{k1}|$, let 
\begin{equation}\hat{S}_{k1}^{*}=[n(n-1)]^{-1}\sum_{i \neq j} |Y_i-Y_j||Z_{i,k}-Z_{j,k}|, \nonumber \end{equation}
then by equation (B.1) in \cite{Lietal2012}: 
\begin{equation}P(|\hat{S}_{k1}-S_{k1}| > Cn^{-\kappa}) \leq P(|\hat{S}_{k1}^{*}-S_{k1}| > Cn^{-\kappa}).\label{B.1} \end{equation}
We also have the following decomposition: 
\begin{equation}|\hat{S}_{k1}^{*}-S_{k1}| \leq |\hat{S}_{k1}^{*}-E(\hat{S}_{k1}^{*})|+|E(\hat{S}_{k1}^{*})-S_{k1}|\label{bias}. \end{equation} Observe that $\hat{S}_{k1}^{*}$ is a $U$-statistic, and is a biased estimate of $S_{k1}$ due to temporal dependence. By condition 3.4, we can control this bias, and we have $|E(\hat{S}_{k1}^{*}-S_{k1})|=O(n^{-\frac{1}{2}})$ by \cite{Yoshihara1976}. Obtaining a bound on $P(|\hat{S}_{k1}^{*}-S_{k1}| > Cn^{-\kappa})$ is difficult in a time series setting. \cite{Borisov} and \cite{Han2016} introduced exponential inequalities for $U$-statistics in a time series setting under uniform mixing type conditions, in addition to restrictions on the kernel function. These restrictions are often too strict and rule out most commonly used time series. For example, even AR(1) processes where the innovations have unbounded support are not uniform mixing (see example 14.8 in \cite{Davidson94}). 

As a result, we will instead rely on Nagaev and Rosenthal type inequalities \citep{WuandWu2016,Liuetal2013} to obtain our bounds. We first show the bounds obtained by using Nagaev inequalities, and then we show the results obtained using Rosenthal type inequalities. Let $\psi_i=(e_i,\bm{\eta}_i)$ and $H_{i,j}=|Y_i-Y_j||Z_{i,k}-Z_{j,k}|$. We have
\begin{equation} H_{i,j}=f(\ldots,\psi_0,\ldots,\psi_{\max(i,j)})\textrm{ and }\hat{S}_{k1}^{*}=2[n(n-1)]^{-1}\sum_{l=1}^{n-1}\sum_{i=1}^{n-l} H_{i,i+l}. \end{equation}
We can then write:
\begin{align} &P(|\sum_{l=1}^{n-1}\sum_{i=1}^{n-l} (H_{i,i+l}-E(H_{i,i+l})) | > Cn^{2-\kappa})\\
\leq\quad & \sum_{l=1}^{n-1}P(|\sum_{i=1}^{n-l} (H_{i,i+l}-E(H_{i,i+l})) | > Cn^{1-\kappa}) \label{decomp}.\end{align}
Note that for any fixed $l$, $\left\{H_{i,i+l}\right\}_{i \in \mathcal{Z}}$ is a Bernoulli shift process, and we can compute the cumulative functional dependence measure as:
\begin{align}
&\quad\sum_{i=m}^{\infty} |||Y_i-Y_{i+l}||Z_{i,k}-Z_{i+l,k}|-|Y_i^{*}-Y_{i+l}^{*}||Z_{i,k}^{*}-Z_{i+l,k}^{*}|||_{\tau}\nonumber\\
 \leq &\quad \sum_{i=m}^{\infty} ||Y_i-Y_{i+l}||_{r}|||Z_{i,k}-Z_{i+l,k}|-|Z_{i,k}^{*}-Z_{i+l,k}^{*}|||_{q}\nonumber \\ 
&+ \sum_{i=m}^{\infty} ||Z_{i,k}^{*}-Z_{i+l,k}^{*}||_{q}|||Y_i-Y_{i+l}|-|Y_i^{*}-Y_{i+l}^{*}|||_{r}\nonumber \\
 \leq&\quad \sum_{i=m}^{\infty} ||Y_i-Y_{i+l}||_{r}|||Z_{i,k}-Z_{i,k}^{*}|+|Z_{i+l,k}-Z_{i+l,k}^{*}|||_{q}\nonumber \\ 
&+ \sum_{i=m}^{\infty} ||Z_{i,k}^{*}-Z_{i+l,k}^{*}||_{q}|||Y_i-Y_i^{*}|+|Y_{i+l}-Y_{i+l}^{*}|||_{r}\nonumber \\
\leq& \quad2\Delta_{0,q}(\bm{y})\Phi_{m,r}(\bm{x})+2\Delta_{m,q}(\bm{y})\Phi_{0,r}(\bm{x})= O(m^{-\alpha}). \label{cume}
\end{align}
The last inequality holds since $||Z_{ik}||_{r} \leq \Phi_{0,r}(\bm{x})$, by section 2 in \cite{WuandWu2016}. Therefore,
\begin{equation}\sup_m (m+1)^{\alpha}\sum_{i=m}^{\infty} |||Y_i-Y_{i+l}||Z_{i,k}-Z_{i+l,k}|-|Y_i^{*}-Y_{i+l}^{*}||Z_{i,k}^{*}-Z_{i+l,k}^{*}|||_{\tau} \leq 4K_{z,r}K_{y,q}.\end{equation}
Using the above result, and theorem 2 in \cite{WuandWu2016}, we obtain:
\begin{equation} P(|\sum_{i=1}^{n-l} (H_{i,i+l}-E(H_{i,i+l})) | > Cn^{1-\kappa}) \leq O(\frac{n^{\iota} K_{z,r}^{\tau}K_{y,q}^{\tau}}{n^{\tau-\tau\kappa}})+ O(\exp(-\frac{n^{1-2\kappa}}{K_{z,r}^2K_{y,q}^2}))\label{bound1}.\end{equation}
Using condition 3.4 along with (\ref{B.1}),(\ref{bias}),(\ref{decomp}), and ({\ref{bound1}}), we obtain:
\begin{equation}P(|\hat{S}_{k1}-S_{k1}| > Cn^{-\kappa}) \leq O(n\frac{n^{\iota} K_{z,r}^{\tau}K_{y,q}^{\tau}}{n^{\tau-\tau\kappa}})+ O(n\exp(-\frac{n^{1-2\kappa}}{K_{z,r}^2K_{y,q}^2})).
\label{S1}\end{equation}
Next let $\hat{S}_{k2}=\hat{S}_{k2,1}\hat{S}_{k2,2}$, where $\hat{S}_{k2,1}=n^{-2}\sum_{j=1}^{n}\sum_{i=1}^{n}|Y_i-Y_j|$ and 
$\hat{S}_{k2,2}=n^{-2}\sum_{j=1}^{n}\sum_{i=1}^{n}|Z_i-Z_j|$. Using this representation we obtain:
\begin{align}
P(|\hat{S}_{k2}-S_{k2}|>Cn^{-\kappa}) &\leq P(|(\hat{S}_{k2,1}-S_{k2,1})S_{k2,2}|>Cn^{-\kappa}) \nonumber \\
&+P(|(\hat{S}_{k2,2}-S_{k2,2})S_{k2,1}|>Cn^{-\kappa}) \nonumber\\
&+P(|(\hat{S}_{k2,1}-S_{k2,1})(\hat{S}_{k2,2}-S_{k2,2})|>Cn^{-\kappa}).\label{Sk2decomp}
\end{align}
Using the same methods as used for $\hat{S}_{k1}$, we obtain:
\begin{align} 
P(|\hat{S}_{k2}-S_{k2}|>Cn^{-\kappa}) & \leq O(n\frac{n^{\zeta} K_{z,r}^{r}}{n^{r-r\kappa}})+ O(n\exp(-\frac{n^{1-2\kappa}}{K_{z,r}^2}))\nonumber\\
&+O(n\frac{n^{\varrho} K_{y,q}^{q}}{n^{q-q\kappa}})+ O(n\exp(-\frac{n^{1-2\kappa}}{K_{y,q}^2})) \label{S2}
.\end{align}
We now proceed to $\hat{S}_{k3}$. As in \cite{Lietal2012}, we define:
\begin{align}
\hat{S}_{k3}^{*}=[n(n-1)(n-2)]^{-1}&\sum_{i<j<l} [|Z_{ik}-Z_{jk}||Y_{j}-Y_{l}|+|Z_{ik}-Z_{lk}||Y_{j}-Y_{l}| \nonumber\\
&+|Z_{ik}-Z_{jk}||Y_{i}-Y_{l}|+|Z_{lk}-Z_{jk}||Y_{i}-Y_{l}|\nonumber \\
&+|Z_{lk}-Z_{jk}||Y_{i}-Y_{j}|+|Z_{lk}-Z_{ik}||Y_{i}-Y_{j}|].
\label{Sk3o}\end{align}
Note that $\hat{S}_{k3}^{*}$ is a $U$-statistic.  Using condition 3.4 and \cite{Yoshihara1976}, we can control its bias: $|E(\hat{S}_{k3}^{*}-S_{k3})|=O(n^{-\frac{1}{2}})$. By equation (A.15) in \cite{Lietal2012}:
\begin{align} P(|\hat{S}_{k3}-S_{k3}|>Cn^{-\kappa})& \leq P(|\hat{S}_{k3}^{*}-S_{k3}|>Cn^{-\kappa})\label{Sk3}\\
&+P(|\hat{S}_{k1}^{*}-S_{k1}|>Cn^{-\kappa})\label{Sk11}.\end{align}
We have already dealt with (\ref{Sk11}), so we will proceed to (\ref{Sk3}). It suffices to deal with the first term in (\ref{Sk3o}), since the rest can be bounded similarly. Let $H_{i,j,l}=|Z_{ik}-Z_{jk}||Y_{j}-Y_{l}|=f(\ldots,\psi_0,\ldots,\psi_{\max(i,j,l)})$. We can then represent
\begin{align}
\sum_{i<j<l} |Z_{ik}-Z_{jk}||Y_{j}-Y_{l}|=\sum_{l=1}^{n-2}\sum_{j=1}^{n-l-1}\sum_{i=1}^{n-j-l}H_{i,i+j,i+j+l}
.\end{align}
Note that for fixed $j,l$, $\left\{H_{i,i+j,i+j+l} \right\}_{i \in \mathcal{Z}}$ is a Bernoulli shift process, whose cumulative functional dependence measure is the same as (\ref{cume}). We can then write:
\begin{align} P(|\sum_{l=1}^{n-2}\sum_{j=1}^{n-l-1}\sum_{i=1}^{n-j-l}[H_{i,i+j,i+j+l}-E(H_{i,i+j,i+j+l})] | > Cn^{3-\kappa}) \nonumber\\
\leq \sum_{l=1}^{n-2}\sum_{j=1}^{n-l-1}P(|\sum_{i=1}^{n-j-l} [H_{i,i+j,i+j+l}-E(H_{i,i+j,i+j+l})] | > Cn^{1-\kappa}). \label{decomp2}\end{align}
Using condition 3.4, along with (\ref{S1}),(\ref{Sk3o}),(\ref{Sk3}),(\ref{Sk11}),(\ref{decomp2}), and theorem 2 in \cite{WuandWu2016}, we obtain:
\begin{equation}P(|\hat{S}_{k3}-S_{k3}| > Cn^{-\kappa}) \leq O(n^{2}\frac{n^{\iota} K_{z,r}^{\tau}K_{y,q}^{\tau}}{n^{\tau-\tau\kappa}})+ O(n^2\exp(-\frac{n^{1-2\kappa}}{K_{z,r}^2K_{y,q}^2}))
\label{S3}.\end{equation}
This gives us a bound for (\ref{triple2}). For (\ref{triple3}): $|\hat{T}_2^{-1}-T_2^{-1}|=|\frac{\hat{T}_2-T_2}{T_2\hat{T_2}}|$ and $T_2$ is finite by condition 3.4. Using this, we obtain:
\begin{align}
P(|\hat{T}_2^{-1}-T_2^{-1}| > Cn^{-\kappa}) & \leq P(|\hat{T}_2-T_2| > |\hat{T_2}|Cn^{-\kappa}) \nonumber\\
&\leq P(|\hat{T}_2-T_2| > CMn^{-\kappa})+ P(|\hat{T}_2| < M) \label{denom}
.\end{align}  

We will deal with the first term in (\ref{denom}) and the second term can be handled similarly. Using the definition of $\hat{T}_2,T_2$ and the decomposition we used in (\ref{Sk2decomp}), it suffices to analyze
\begin{align} &P(|\widehat{dcov}(Y_t,Y_t)-dcov(Y_t,Y_t)| > Cn^{-\kappa}) \label{dcY}\\
&\textrm{and }P(|\widehat{dcov}(Z_{t-1,k},Z_{t-1,k})-dcov(Z_{t-1,k},Z_{t-1,k})|>Cn^{-\kappa})\label{dcZ}
.\end{align}
For (\ref{dcY}) and (\ref{dcZ}), note that for $a>0,b>0$ we have $|\sqrt{a}-\sqrt{b}| =\frac{|a-b|}{\sqrt{a}+\sqrt{b}} < \frac{|a-b|}{\sqrt{b}} $. Using this, along with (\ref{denom}) and the methods  used to bound $\hat{T}_1$, we obtain:
\begin{align} 
P(|\hat{T}_{2}^{-1}-T_{2}^{-1}|>Cn^{-\kappa}) & \leq O(n^2\frac{n^{\zeta} K_{z,r}^{r}}{n^{r/2-r\kappa/2}})+ O(n^2\exp(-\frac{n^{1-2\kappa}}{(K_{z,r})^2}))\nonumber\\
&+O(n^2\frac{n^{\varrho} K_{y,q}^{q}}{n^{q/2-q\kappa/2}})+ O(n^2\exp(-\frac{n^{1-2\kappa}}{(K_{y,q})^2})).\label{T2}
\end{align}
By (\ref{S1}),(\ref{S2}),(\ref{S3}),(\ref{T2}), we obtain:
\begin{align} 
P(|\frac{\hat{T}_1}{\hat{T}_2}-\frac{T_1}{T_2}|> cn^{-\kappa}) & \leq O(n^2\frac{n^{\zeta} K_{z,r}^{r}}{n^{r/2-r\kappa/2}})+ O(n^2\exp(-\frac{n^{1-2\kappa}}{K_{z,r}^2}))\nonumber\\
&+O(n^2\frac{n^{\varrho} K_{y,q}^{q}}{n^{q/2-q\kappa/2}})+ O(n^2\exp(-\frac{n^{1-2\kappa}}{K_{y,q}^2}))\nonumber\\
&+O(n^{2}\frac{n^{\iota} K_{z,r}^{\tau}K_{y,q}^{\tau}}{n^{\tau-\tau\kappa}})+ O(n^2\exp(-\frac{n^{1-2\kappa}}{K_{z,r}^2K_{y,q}^2}))
.\label{NagaevA}\end{align}
The other terms in (\ref{pddef}) deal with the conditioning vectors $C_j$, and we need to account for the maximum dimension of the conditioning vectors $\max_{j}[dim(C_j)]=t_n$. This comes into effect when computing the cumulative functional dependence measure. Recall that $C_{k+(h-1)*m_n}=\mathcal{S}_{k,h}$, and for analyzing the cumulative functional dependence measure, we define
\begin{equation} \mathcal{S}_{k,h}(i)=\left\{Y_{i-1},\ldots,Y_{i-h},X_{i-1,k},\ldots,X_{i-h+1,k} \right\}, \end{equation} 
as the conditional vector of the $h^{th}$ lag of series $k$ at time $i$. Additionally recall that $|\bm{a}|_p$ stands for the Euclidean norm of $ \bm{a} \in \mathcal{R}^p$. Assume $dim(S_{k,h})=t_n$ and $q=r$, we therefore have:
\begin{align}
&\sum_{i=m}^{\infty} \biggl(|||\mathcal{S}_{k,h}(i)-\mathcal{S}_{k,h}(i+j)|_{t_n}|\mathcal{S}_{k,h}(i)-\mathcal{S}_{k,h}(i+j)|_{t_n}\\
&-|\mathcal{S}_{k,h}^{*}(i)-\mathcal{S}_{k,h}^{*}(i+j)|_{t_n}|\mathcal{S}_{k,h}^{*}(i)-\mathcal{S}_{k,h}^{*}(i+j)|_{t_n}||_{q/2}\biggr)\nonumber\\
& \leq \sum_{i=m}^{\infty} |||\mathcal{S}_{k,h}(i)-\mathcal{S}_{k,h}(i+j)|_{t_n}||_{q}|||\mathcal{S}_{k,h}(i)-\mathcal{S}_{k,h}(i+j)|_{t_n}-|\mathcal{S}_{k,h}^{*}(i)-\mathcal{S}_{k,h}^{*}(i+j)|_{t_n}||_{q}\nonumber \\ 
&+ \sum_{i=m}^{\infty} |||\mathcal{S}_{k,h}^{*}(i)-\mathcal{S}_{k,h}^{*}(i+j)|_{t_n}||_{q}|||\mathcal{S}_{k,h}(i)-\mathcal{S}_{k,h}(i+j)|_{t_n}-|\mathcal{S}_{k,h}^{*}(i)-\mathcal{S}_{k,h}^{*}(i+j)|_{t_n}||_{q}\nonumber \\
&  \leq \sum_{i=m}^{\infty} |||\mathcal{S}_{k,h}(i)-\mathcal{S}_{k,h}(i+j)|_{t_n}||_{q}|||\mathcal{S}_{k,h}(i)-\mathcal{S}_{k,h}^{*}(i)|_{t_n}+|\mathcal{S}_{k,h}(i+j)-\mathcal{S}_{k,h}^{*}(i+j)|_{t_n}||_{q}\nonumber \\ 
&+  \sum_{i=m}^{\infty} |||\mathcal{S}_{k,h}^{*}(i)-\mathcal{S}_{k,h}^{*}(i+j)|_{t_n}||_{q}|||\mathcal{S}_{k,h}(i)-\mathcal{S}_{k,h}^{*}(i)|_{t_n}+|\mathcal{S}_{k,h}(i+j)-\mathcal{S}_{k,h}^{*}(i+j)|_{t_n}||_{q}\nonumber \\
&\leq t_n(\Delta_{0,q}(\bm{y})+\Phi_{m,q}(\bm{x}))^{2} \nonumber
.\end{align}

To explain the last inequality, we analyze the term: 
\begin{align}\sum_{i=m}^{\infty}|||\mathcal{S}_{k,h}(i)-\mathcal{S}_{k,h}^{*}(i)|_{t_n}||_{q}&=\sum_{i=m}^{\infty}|||\mathcal{S}_{k,h}(i)-\mathcal{S}_{k,h}^{*}(i)|_{t_n}^2||_{q/2}^{1/2}\nonumber \\
& \leq (t_n/2)^{1/2}(\Delta_{0,q}(\bm{y})+\Phi_{0,q}(\bm{x}))
.\end{align}
Where the last inequality follows from Minkowski's inequality and the definition of $S_{k,h}(i)$. Using this, the rest of the terms in (\ref{pddef}) can be handled as done previously.
\newline\newline
We now show the bounds obtained using a Rosenthal type inequality. We follow the same steps as previously, and it suffices to consider (\ref{Sk3}). As before we  focus on the following term
\begin{align}
\sum_{i<j<l} [|Z_{ik}-Z_{jk}||Y_{j}-Y_{l}|]=\sum_{l=1}^{n-2}\sum_{j=1}^{n-l-1}\sum_{i=1}^{n-j-l}H_{i,i+j,i+j+l}
.\end{align}
Let $Q=[(n-1)(n-2)]^{-1}\sum_{l=1}^{n-2}\sum_{j=1}^{n-l-1}\sum_{i=1}^{n-j-l}H_{i,i+j,i+j+l}$. Then by Markov's inequality we obtain:
\begin{equation}
P(|Q-E(Q)|> cn^{1-\kappa}) \leq \frac{||Q-E(Q)||_{\tau}^{\tau}}{n^{\tau-\tau\kappa}}.
\end{equation}
Then using Minkowski's inequality, we obtain:
\begin{equation}||Q-E(Q)||_{\tau} \leq ||\sum_{i=1}^{n-2}H_{i,i+1,i+2}-E(H_{i,i+1,i+2})||_{\tau}
.\label{Jensen2}\end{equation}
As we stated previously, for fixed $j,l$, $\left\{H_{i,i+j,i+j+l} \right\}_{i \in \mathcal{Z}}$ is a Bernoulli shift process whose cumulative functional dependence measure is the same as (\ref{cume}). By theorem 1 in \cite{Liuetal2013}, we have:
\begin{equation}||\sum_{i=1}^{n-2}H_{i,i+1,i+2}-E(H_{i,i+1,i+2})||_{\tau} \leq O(K_{z,r}K_{y,q}n^{\frac{1}{2}}). \end{equation}
Combining the above with (\ref{Jensen2}), we obtain:
\begin{equation} P(|Q-E(Q)|> cn^{1-\kappa}) \leq  O(\frac{K_{z,r}^{\tau}K_{y,q}^{\tau}n^{\frac{\tau}{2}}}{n^{\tau-\tau\kappa}}). \end{equation}
By repeating the same techniques we obtain:
\begin{equation} P(|\hat{\omega}_k -\omega_k|>c_2 n^{-\kappa}) \leq O(\frac{K_{y,q}^{q}n^{\frac{q}{4}}}{n^{q/2-q\kappa/2}})+O(\frac{K_{z,r}^{\tau}K_{y,q}^{\tau}n^{\frac{\tau}{2}}}{n^{\tau-\tau\kappa}})+O(\frac{K_{z,r}^{r}n^{\frac{r}{4}}}{n^{r/2-r\kappa/2}}) 
\label{RosenthalA}.\end{equation}
For simplicity we assume $r=q$, and we now compare the above result to (\ref{NagaevA}), which was obtained using Nagaev type inequalities. Note that when $q=r$ the above bound is of the order $O(n^{r/4-r\kappa/2})$. Using Nagaev type inequalities leads to the bound at most  $O(n^{r/2-r\kappa/2-3})$. Therefore, when $r<12$, (\ref{RosenthalA}) provides a better bound. When $r>12$, the comparison depends on the values of $\varrho,\iota,\zeta$ which are related to the dependence of the covariate and response processes. Applying the union bound gives us the desired result.
\newline\newline
For part (iv), let $\mathcal{A}_n=\{\max_{k \in M_*}|\hat{\rho}_k-\rho_k|\leq \frac{c_1 n^{-\kappa}}{2}\}$. On the set $\mathcal{A}_n$, by condition 3.1, we have: 
\begin{equation} |\hat{\rho}_k| \geq |\rho_k|-|\hat{\rho}_k - \rho_k| \geq c_1 n^{-\kappa}/2, \textrm{  }\forall k \in M_*.\end{equation}
Hence by our choice of $\gamma_n$, we obtain $P\left(\mathcal{M}_*\subset \hat{\mathcal{M}}_{\gamma_n} \right) > P(\mathcal{A}_n)$. By applying part (i), the result follows.
\newline\newline
For part(i), we first define the \emph{predictive dependence measure} introduced by \cite{Wu2005}. The predictive dependence measure for a univariate process and multivariate processes is defined respectively as:
\begin{align}\theta_{q}(y_i)=||\textrm{E}\left(y_i|\mathcal{F}_{0}\right)-\textrm{E}\left(y_i|\mathcal{F}_{-1}\right)||_q,\nonumber\\
\theta_{q}(Z_{ij})=||\textrm{E}\left(Z_{ij}|\mathcal{H}_{0}\right)-\textrm{E}\left(Z_{ij}|\mathcal{H}_{-1}\right)||_q .\end{align}
With the cumulative predictive dependence measures defined as:
\begin{equation}\Theta_{0,q}(\bm{x})=\max_{j \leq p_n}\sum_{i=0}^{\infty}\delta_{q}(Z_{ij}), \textrm{ and } \Theta_{0,q}(\bm{\epsilon})=\sum_{i=0}^{\infty}\delta_{q}(\epsilon_i).\end{equation} 
We follow the steps of the proof of part (iii). For $|\widehat{dcor}^2(Y_t,Z_{t-1,k})-{dcor}^2(Y_t,Z_{t-1,k})|$, it suffices to provide a bound for (\ref{decomp2}). Note that for fixed $j,l$, we have:
\begin{align}
\sup_{q \geq 4}q^{-(\tilde{\alpha}_z+\tilde{\alpha}_y)}\sum_{i=1}^{\infty}\theta_{q}(H_{i,i+j,i+j+l}) &\leq \sup_{q \geq 4}q^{-(\tilde{\alpha}_z+\tilde{\alpha}_y)}\sum_{i=1}^{\infty}\delta_{q}(H_{i,i+j,i+j+l}) \nonumber
\\ &\leq \sup_{q \geq 4}q^{-(\tilde{\alpha}_z+\tilde{\alpha}_y)}\Delta_{0,q}(\bm{y}) \Phi_{0,q}(\bm{x}) < \infty 
,\end{align}
where the first inequality follows from theorem 1 in \cite{Wu2005}, and the last inequality follows from condition 3.3. Using the above we have by theorem 3 in \cite{WuandWu2016}:
\begin{equation}
(\ref{decomp2}) \leq O\left(n^2\exp\left(-\frac{n^{1/2-\kappa}}{\upsilon_z\upsilon_{y}}\right)^{\tilde{\psi}}\right)
.\end{equation}
We now provide a bound for (\ref{dcY}) in a similar way. Let $S_{i,j,l}=|Y_{i}-Y_{j}||Y_{j}-Y_{l}| =f_1(\ldots,e_0,\ldots,e_{\max(i,j,l)})$. We then have:
\begin{align}
\sup_{q \geq 4}q^{-2\tilde{\alpha}_y}\sum_{i=1}^{\infty}\theta_{q}(S_{i,i+j,i+j+l}) &\leq \sup_{q \geq 4}q^{-2\tilde{\alpha}_y}\sum_{i=1}^{\infty}\delta_{q}(S_{i,i+j,i+j+l}) \nonumber
\\ &\leq \sup_{q \geq 4}q^{-2\tilde{\alpha}_y}\Delta_{0,q}^2(\bm{y})< \infty 
.\end{align}
Then by theorem 3 in \cite{WuandWu2016}:
\begin{align}
\sum_{l=1}^{n-2}\sum_{j=1}^{n-l-1}P(|\sum_{i=1}^{n-j-l} (S_{i,i+j,i+j+l}-E(S_{i,i+j,i+j+l})) |& > Cn^{1-\kappa})\nonumber\\
\leq O\left(n^2\exp\left(-\frac{n^{1/2-\kappa}}{\upsilon_y^2}\right)^{\tilde{\alpha}}\right)
.\end{align}
A similar result holds for (\ref{dcZ}). Following the steps in the proof of part (iii), and using the results above we obtain:
\begin{align}
P(\max_{j \leq p_n}|\hat{\omega}_k -\omega_k|>c_2 n^{-\kappa}) &\leq p_n\bigg[O(n^2\exp\left(-\frac{n^{1/2-\kappa}}{\upsilon_y^2}\right)^{\tilde{\alpha}})\nonumber\\
&+ O(n^2\exp\left(-\frac{n^{1/2-\kappa}}{\upsilon_z\upsilon_{y}}\right)^{\tilde{\psi}})\nonumber\\
&+ O(n^2\exp\left(-\frac{n^{1/2-\kappa}}{\upsilon_z^2}\right)^{\tilde{\varphi}})\bigg].\nonumber
\end{align}
The proof for part (ii) is similar to the proof for part (iv) and we omit its details.
\end{proof}

\begin{proof} [Proof of Theorem 2]
  $ $\newline

For simplicity we only prove part (i), and the proof for part (iii) follows similarly. Let $\tilde{\bm{\omega}}=(\tilde{\omega}_1,\ldots,\tilde{\omega}_{p_n})$, where $\tilde{\omega}_k=\widehat{pdcor}(Y_t,Z_{t-1,k};\hat{C}_k)$. We will work on the following set,
\begin{align*} 
\mathcal{A}_n=\{\max_{k \leq p_n}|\tilde{\omega}_k -\omega_k| \leq \frac{c_1}{2} n^{-\kappa} \} 
.\end{align*}
The main difference in the proof for this procedure vs. PDC-SIS lies in the randomness which results from estimating the conditional sets at each lag level. We claim that on the set $\mathcal{A}_n$,  $\hat{\bm{C}}=\bm{C}$. To see this, note that on the first lag level: $\max_{k \leq m_n}|\tilde{\omega}_k -\omega_k| \leq \frac{c_1}{2} n^{-\kappa}$, 
which implies $\hat{\mathcal{U}}_{1}^{\lambda_n}=\mathcal{U}_{1}^{\lambda_n}$. Now due to $\hat{\mathcal{U}}_{1}^{\lambda_n}=\mathcal{U}_{1}^{\lambda_n}$, we have $\hat{C}_j=C_j$ for $k \in {m_n+1,\ldots,2m_n}$, which implies $\tilde{\omega}_k=\hat{\omega}_k$ for $k\in {m_n+1,\ldots,2m_n}$. Continuing this argument we see that on the set $\mathcal{A}_n$ we have $\hat{\bm{C}}=\bm{C}$, and therefore $\tilde{\bm{\omega}}=\hat{\bm{\omega}}$. The result then follows from the results in theorem 1. 
\end{proof}

\section*{Supplement C: Tables for Section 5.1} \label{sec:SuppB}

Tables \ref{table1}-\ref{table4} provide more detailed results of the simulations in section 5.1. As stated in our main paper, tables \ref{table1}-\ref{table4} report the median minimum model size needed to include all the relevant predictors, as well as the median rank of the significant covariates for each procedure.
\begin{table}[]
\centering
\caption{Model 1}
\label{table1}
\begin{tabular}{|c|c|c|c|c|c|c|c|}
\hline
\multicolumn{8}{|c|}{Gaussian, $p_n=1500$}                                                                                                                                           \\ \hline
     & MMS & $X_{t-1,1}$ & $X_{t-1,2}$ & \multicolumn{1}{l|}{$X_{t-1,3}$} & \multicolumn{1}{l|}{$X_{t-1,4}$} & \multicolumn{1}{l|}{$X_{t-1,5}$} & \multicolumn{1}{l|}{$X_{t-1,6}$} \\ \hline
PDC-SIS  & 7   & 6           & 3           & 2                                & 2                                & 3                                & 5                                \\ \hline
DC-SIS   & 11  & 7           & 3.5         & 2                                & 2                                & 3                                & 5.5                              \\ \hline
NIS  & 11  & 6           & 3           & 2                                & 2                                & 3                                & 6                                \\ \hline
SIS  & 10  & 6           & 3           & 2                                & 2                                & 3                                & 6                                \\ \hline
GLSS & 6   & 5           & 3           & 2                                & 2                                & 3                                & 5                                \\ \hline
\end{tabular}
\begin{tabular}{|c|c|c|c|c|c|c|c|}
\hline
\multicolumn{8}{|c|}{Gaussian, $p_n=4500$}                                                                                                                                          \\ \hline
     & MMS & $X_{t-1,1}$ & $X_{t-1,2}$ & \multicolumn{1}{l|}{$X_{t-1,3}$} & \multicolumn{1}{l|}{$X_{t-1,4}$} & \multicolumn{1}{l|}{$X_{t-1,5}$} & \multicolumn{1}{l|}{$X_{t-1,6}$} \\ \hline
PDC-SIS  & 11  & 5           & 3           & 3                                & 3                                & 3                                & 5                                \\ \hline
DC-SIS   & 19  & 6           & 3           & 3                                & 3                                & 3                                & 6                                \\ \hline
NIS  & 16  & 6           & 3           & 3                                & 3                                & 3                                & 6                                \\ \hline
SIS  & 13  & 5           & 3           & 2.5                              & 3                                & 3                                & 6                                \\ \hline
GLSS & 6   & 5           & 3           & 2                                & 2                                & 3                                & 5                                \\ \hline
\end{tabular}
\begin{tabular}{|c|c|c|c|c|c|c|c|}
\hline
\multicolumn{8}{|c|}{$t_5$, $p_n=1500$}                                                                                                                                               \\ \hline
     & MMS  & $X_{t-1,1}$ & $X_{t-1,2}$ & \multicolumn{1}{l|}{$X_{t-1,3}$} & \multicolumn{1}{l|}{$X_{t-1,4}$} & \multicolumn{1}{l|}{$X_{t-1,5}$} & \multicolumn{1}{l|}{$X_{t-1,6}$} \\ \hline
PDC-SIS  & 13   & 5           & 3           & 3                                & 3                                & 3                                & 5                                \\ \hline
DC-SIS   & 20   & 6           & 4           & 3                                & 3                                & 3                                & 6                                \\ \hline
NIS  & 33   & 7           & 4           & 3                                & 3                                & 3                                & 6                                \\ \hline
SIS  & 21.5 & 6           & 3           & 3                                & 3                                & 3                                & 5                                \\ \hline
GLSS & 6    & 5           & 3           & 2                                & 2                                & 3                                & 5                                \\ \hline
\end{tabular}
\begin{tabular}{|c|c|c|c|c|c|c|c|}
\hline
\multicolumn{8}{|c|}{$t_5$, $p_n=4500$}                                                                                                                                              \\ \hline
     & MMS  & $X_{t-1,1}$ & $X_{t-1,2}$ & \multicolumn{1}{l|}{$X_{t-1,3}$} & \multicolumn{1}{l|}{$X_{t-1,4}$} & \multicolumn{1}{l|}{$X_{t-1,5}$} & \multicolumn{1}{l|}{$X_{t-1,6}$} \\ \hline
PDC-SIS  & 36.5 & 7           & 4           & 2                                & 2                                & 3                                & 5                                \\ \hline
DC-SIS   & 68   & 10.5        & 4           & 2                                & 3                                & 3                                & 7                                \\ \hline
NIS  & 114  & 16.5        & 4           & 2                                & 3                                & 4                                & 9                                \\ \hline
SIS  & 66.5 & 10.5        & 4           & 3                                & 3                                & 4                                & 7                                \\ \hline
GLSS & 7    & 5           & 3           & 2                                & 2                                & 3                                & 5                                \\ \hline
\end{tabular}
\end{table}

\begin{table}[]
\centering
\caption{Model 2}
\label{table2}
\begin{tabular}{|c|c|c|c|c|c|}
\hline
\multicolumn{6}{|c|}{Gaussian $p_n$=1500}                                                                      \\ \hline
     & MMS   & $X_{t-1,1}$ & $X_{t-2,1}$ & \multicolumn{1}{l|}{$X_{t-1,2}$} & \multicolumn{1}{l|}{$X_{t-2,2}$} \\ \hline
PDC-SIS  & 61    & 1           & 40.5        & 2                                & 5                                \\ \hline
DC-SIS   & 488   & 1           & 488         & 2                                & 3                                \\ \hline
NIS  & 488   & 1           & 488         & 2                                & 3                                \\ \hline
SIS  & 343.5 & 1           & 341.5       & 2                                & 3                                \\ \hline
GLSS & 179.5 & 1           & 160.5       & 2                                & 6.5                              \\ \hline
\end{tabular}
\begin{tabular}{|c|c|c|c|c|c|}
\hline
\multicolumn{6}{|c|}{Gaussian $p_n$=4500}                                                                     \\ \hline
     & MMS  & $X_{t-1,1}$ & $X_{t-2,1}$ & \multicolumn{1}{l|}{$X_{t-1,2}$} & \multicolumn{1}{l|}{$X_{t-2,2}$} \\ \hline
PDC-SIS  & 149  & 1           & 141         & 2                                & 4                                \\ \hline
DC-SIS   & 1051 & 1           & 1051        & 2                                & 3                                \\ \hline
NIS  & 861  & 1           & 861         & 2                                & 3                                \\ \hline
SIS  & 722  & 1           & 722         & 2                                & 3                                \\ \hline
GLSS & 592  & 1           & 412.5       & 2                                & 8                                \\ \hline
\end{tabular}
\begin{tabular}{|c|c|c|c|c|c|}
\hline
\multicolumn{6}{|c|}{$t_5$ $p_n$=1500}                                                                         \\ \hline
     & MMS   & $X_{t-1,1}$ & $X_{t-2,1}$ & \multicolumn{1}{l|}{$X_{t-1,2}$} & \multicolumn{1}{l|}{$X_{t-2,2}$} \\ \hline
PDC-SIS  & 79.5  & 1           & 57.5     & 2                                & 5                                \\ \hline
DC-SIS   & 408.5 & 1           & 408.5       & 2                                & 3                                \\ \hline
NIS  & 513.5 & 1           & 492         & 2                                & 4                                \\ \hline
SIS  & 447   & 1           & 440         & 2                                & 4                                \\ \hline
GLSS & 450.5 & 1           & 330.5       & 2                                & 22                               \\ \hline
\end{tabular}
\begin{tabular}{|c|c|c|c|c|c|}
\hline
\multicolumn{6}{|c|}{$t_5$ $p_n$=4500}                                                                          \\ \hline
     & MMS    & $X_{t-1,1}$ & $X_{t-2,1}$ & \multicolumn{1}{l|}{$X_{t-1,2}$} & \multicolumn{1}{l|}{$X_{t-2,2}$} \\ \hline
PDC-SIS  & 275.5  & 1           & 239.5       & 2                                & 5                                \\ \hline
DC-SIS   & 951.5  & 1           & 951.5       & 2                                & 3                                \\ \hline
NIS  & 1100.5 & 1           & 984         & 2                                & 4                                \\ \hline
SIS  & 905    & 1           & 859.5       & 2                                & 3                                \\ \hline
GLSS & 1386.5 & 1           & 995         & 2                                & 18.5                             \\ \hline
\end{tabular}
\end{table}

\begin{table}[]
\centering
\caption{Model 3}
\label{table3}
\begin{tabular}{|c|c|c|c|c|c|c|c|}
\hline
\multicolumn{8}{|c|}{Gaussian, $p_n=1500$}                                                                                                                                             \\ \hline
     & MMS   & $X_{t-1,1}$ & $X_{t-2,1}$ & \multicolumn{1}{l|}{$X_{t-1,2}$} & \multicolumn{1}{l|}{$X_{t-2,2}$} & \multicolumn{1}{l|}{$X_{t-1,3}$} & \multicolumn{1}{l|}{$X_{t-1,4}$} \\ \hline
PDC-SIS  & 29    & 2           & 4           & 4                                & 3                                & 7                                & 11                               \\ \hline
DC-SIS   & 112   & 8           & 4.5         & 8                                & 4                                & 19                               & 34.5                             \\ \hline
NIS  & 119.5 & 8           & 4           & 8                                & 3                                & 18                               & 48.5                             \\ \hline
SIS  & 100.5 & 7           & 4           & 7                                & 3                                & 16                               & 42                               \\ \hline
GLSS & 813   & 14.5        & 164         & 535.5                            & 13                               & 2                                & 18                               \\ \hline
\end{tabular}
\begin{tabular}{|c|c|c|c|c|c|c|c|}
\hline
\multicolumn{8}{|c|}{Gaussian, $p_n=4500$}                                                                                                                                             \\ \hline
     & MMS    & $X_{t-1,1}$ & $X_{t-2,1}$ & \multicolumn{1}{l|}{$X_{t-1,2}$} & \multicolumn{1}{l|}{$X_{t-2,2}$} & \multicolumn{1}{l|}{$X_{t-1,3}$} & \multicolumn{1}{l|}{$X_{t-1,4}$} \\ \hline
PDC-SIS  & 78.5   & 3           & 4           & 3.5                              & 2                                & 10                               & 20                               \\ \hline
DC-SIS   & 337    & 15          & 6.5         & 10                               & 3                                & 19                               & 34.5                             \\ \hline
NIS  & 309    & 14          & 6           & 9                                & 3                                & 39                               & 137                              \\ \hline
SIS  & 281    & 11.5        & 5           & 8                                & 2                                & 31                               & 130                              \\ \hline
GLSS & 2325.5 & 30.5        & 364         & 1709.5                           & 36.5                             & 2                                & 73.5                             \\ \hline
\end{tabular}
\begin{tabular}{|c|c|c|c|c|c|c|c|}
\hline
\multicolumn{8}{|c|}{$t_5$, $p_n=1500$}                                                                                                                                               \\ \hline
     & MMS   & $X_{t-1,1}$ & $X_{t-2,1}$ & \multicolumn{1}{l|}{$X_{t-1,2}$} & \multicolumn{1}{l|}{$X_{t-2,2}$} & \multicolumn{1}{l|}{$X_{t-1,3}$} & \multicolumn{1}{l|}{$X_{t-1,4}$} \\ \hline
PDC-SIS  & 43    & 3           & 3.5         & 4                                & 3                                & 6.5                              & 16                               \\ \hline
DC-SIS   & 114   & 8           & 5           & 9                                & 4                                & 16                               & 64.5                             \\ \hline
NIS  & 167   & 9           & 4           & 11                               & 4                                & 15                               & 51                               \\ \hline
SIS  & 166.5 & 8           & 4           & 10                               & 4                                & 18.5                             & 71                               \\ \hline
GLSS & 969.5 & 42          & 202         & 453                              & 60.5                             & 3                                & 44                               \\ \hline
\end{tabular}
\begin{tabular}{|c|c|c|c|c|c|c|c|}
\hline
\multicolumn{8}{|c|}{$t_5$, $p_n=4500$}                                                                                                                                              \\ \hline
     & MMS   & $X_{t-1,1}$ & $X_{t-2,1}$ & \multicolumn{1}{l|}{$X_{t-1,2}$} & \multicolumn{1}{l|}{$X_{t-2,2}$} & \multicolumn{1}{l|}{$X_{t-1,3}$} & \multicolumn{1}{l|}{$X_{t-1,4}$} \\ \hline
PDC-SIS  & 78    & 2           & 5           & 4                                & 3                                & 11                               & 24.5                             \\ \hline
DC-SIS   & 301.5 & 14.5        & 8           & 11.5                             & 4                                & 33                               & 113.5                            \\ \hline
NIS  & 436.5 & 14.5        & 8           & 14                               & 4                                & 33.5                             & 124                              \\ \hline
SIS  & 438   & 13          & 7           & 13                               & 4                                & 33                               & 149.5                            \\ \hline
GLSS & 3008  & 85.5        & 690         & 1362                             & 99.5                             & 9                                & 117.5                            \\ \hline
\end{tabular}
\end{table}
\begin{table}[]
\centering
\caption{Model 4}
\label{table4}
\begin{tabular}{|c|c|c|c|c|c|c|c|}
\hline
\multicolumn{8}{|c|}{Gaussian, $p_n=1500$}                                                                                                                                           \\ \hline
     & MMS   & $X_{t-1,1}$ & $X_{t-2,1}$ & \multicolumn{1}{l|}{$X_{t-1,2}$} & \multicolumn{1}{l|}{$X_{t-2,2}$} & \multicolumn{1}{l|}{$X_{t-1,3}$} & \multicolumn{1}{l|}{$X_{t-2,3}$} \\ \hline
PDC-SIS  & 42    & 5           & 5           & 3                                & 2                                & 20                               & 10                               \\ \hline
DC-SIS   & 306.5 & 114.5       & 53          & 64                               & 22.5                             & 162.5                            & 73                               \\ \hline
NIS  & 275   & 105.5       & 47          & 46                               & 16                               & 149                              & 80                               \\ \hline
SIS  & 234.5 & 95          & 42          & 41                               & 15                               & 129.5                            & 72.5                             \\ \hline
GLSS & 800.5 & 1           & 12          & 5.5                              & 10                               & 552.5                            & 103                              \\ \hline
\end{tabular}
\begin{tabular}{|c|c|c|c|c|c|c|c|}
\hline
\multicolumn{8}{|c|}{Gaussian, $p_n=4500$}                                                                                                                                           \\ \hline
     & MMS   & $X_{t-1,1}$ & $X_{t-2,1}$ & \multicolumn{1}{l|}{$X_{t-1,2}$} & \multicolumn{1}{l|}{$X_{t-2,2}$} & \multicolumn{1}{l|}{$X_{t-1,3}$} & \multicolumn{1}{l|}{$X_{t-2,3}$} \\ \hline
PDC-SIS  & 100.5 & 8           & 6           & 4                                & 2                                & 33                               & 16                               \\ \hline
DC-SIS   & 842.5 & 338         & 144         & 148                              & 53                               & 350                              & 181                              \\ \hline
NIS  & 704   & 255.5       & 104.5       & 119                              & 38                               & 322                              & 158                              \\ \hline
SIS  & 588   & 224         & 95.5        & 103.5                            & 35                               & 307                              & 142                              \\ \hline
GLSS & 2214  & 1           & 29          & 13.5                             & 22                               & 1490.5                           & 291.5                            \\ \hline
\end{tabular}
\begin{tabular}{|c|c|c|c|c|c|c|c|}
\hline
\multicolumn{8}{|c|}{$t_5$, $p_n=1500$}                                                                                                                                              \\ \hline
     & MMS   & $X_{t-1,1}$ & $X_{t-2,1}$ & \multicolumn{1}{l|}{$X_{t-1,2}$} & \multicolumn{1}{l|}{$X_{t-2,2}$} & \multicolumn{1}{l|}{$X_{t-1,3}$} & \multicolumn{1}{l|}{$X_{t-2,3}$} \\ \hline
PDC-SIS  & 51    & 4           & 5           & 5                                & 4                                & 19                               & 9                                \\ \hline
DC-SIS   & 306   & 108.5       & 54.5        & 75                               & 34.5                             & 132                              & 59                               \\ \hline
NIS  & 328   & 90.5        & 39          & 70                               & 27                               & 136                              & 61                               \\ \hline
SIS  & 265   & 79.5        & 33          & 62.5                             & 24.5                             & 133                              & 57                               \\ \hline
GLSS & 891.5 & 3           & 48.5        & 47.5                             & 43                               & 476                              & 162                              \\ \hline
\end{tabular}
\begin{tabular}{|c|c|c|c|c|c|c|c|}
\hline
\multicolumn{8}{|c|}{$t_5$, $p_n=4500$}                                                                                                                                               \\ \hline
     & MMS    & $X_{t-1,1}$ & $X_{t-2,1}$ & \multicolumn{1}{l|}{$X_{t-1,2}$} & \multicolumn{1}{l|}{$X_{t-2,2}$} & \multicolumn{1}{l|}{$X_{t-1,3}$} & \multicolumn{1}{l|}{$X_{t-2,3}$} \\ \hline
PDC-SIS  & 104    & 8           & 8           & 4                                & 3                                & 33                               & 18.5                             \\ \hline
DC-SIS   & 814.5  & 322         & 157         & 155.5                            & 61.5                             & 395                              & 196                              \\ \hline
NIS  & 851.5  & 283         & 139.5       & 144                              & 54.5                             & 418.5                            & 181                              \\ \hline
SIS  & 761    & 249         & 120         & 120.5                            & 46                               & 372                              & 181                              \\ \hline
GLSS & 2843.5 & 5           & 137         & 81                               & 80                               & 1760                             & 554.5                            \\ \hline
\end{tabular}
\end{table}

\clearpage



\end{document}